\documentclass[reprint,
     amsmath,amssymb,aps,superscriptaddress,
]{revtex4-2}

\usepackage{graphicx}  % Include figure files
\usepackage{dcolumn}   % Align table columns on decimal point
\usepackage{bm}        % bold math
\usepackage{graphicx}
\usepackage{epstopdf, epsfig}
\usepackage{amsmath, amssymb}
\usepackage{mathtools}
\usepackage{textcomp}
\usepackage{relsize}
\usepackage{epstopdf}
\usepackage{nth}
\usepackage[]{algorithm2e}
\usepackage{array}
\usepackage{subfigure}
\usepackage{color}
\usepackage{multirow}
\begin{document}

% APS Template Requirement
\preprint{APS/123-QED}

% Title
\title{Drag of a Heated Sphere at Low Reynolds Numbers in the Presence of Buoyancy}

% Author
\author{Swetava Ganguli}
\email{Corresponding Author. EMail: swetava@cs.stanford.edu}
\affiliation{Department of Mechanical Engineering, Stanford University, Stanford, CA 94305}
\affiliation{Department of Computer Science, Stanford University, Stanford, CA 94305}
\author{Sanjiva K. Lele}
\affiliation{Department of Mechanical Engineering, Stanford University, Stanford, CA 94305}
\affiliation{Department of Aeronautics and Astronautics, Stanford University, Stanford, CA 94305}

% Abstract
\begin{abstract}
Fully resolved simulations are used to quantify the effects of heat transfer in the presence of buoyancy on the drag of a spatially fixed heated spherical particle at low Reynolds numbers ($Re$) in the range $10^{-3} \le Re \le 10$ in a variable property fluid, extending the analysis presented in \cite{ganguli2019drag}. The amount of heat addition from the sphere encompasses both, the heating regime where the Boussinesq approximation holds and the regime where it breaks down. The particle is assumed to have a low Biot number which means that the particle is uniformly at the same temperature and has no internal temperature gradients. Scaling buoyancy with inertial and viscous forces yields two related non-dimensional quantities, called \textit{Buoyancy Induced Viscous Reynolds Number} ($Re_{BV}$) and \textit{Buoyancy Induced Inertial Reynolds Number} ($Re_{BI}$). For ideal gases, $Re_{BV}$ is analogous to the Grashof number ($Gr$). No assumptions are made on the magnitude of $Re_{BI}$ (or equivalently $Re_{BV}$). The effects of the orientation of gravity relative to the free-stream velocity are examined. Large deviations in the value of the drag coefficient are observed when the Froude number ($Fr$) decreases and/or the temperature of the sphere increases. Under appropriate constraints on $Re_{BI}$ and $Re$, the total drag on a heated sphere in a low $Re$ flow in the presence of buoyancy (mixed convection) is shown to be, within 10\% error, the linear superposition of the drag computed in two canonical setups: one being the drag on a steadily moving heated sphere in the absence of buoyancy (forced convection) and the other being natural convection. However, the effect of temperature variation on the drag of a sphere in both, forced and natural convection, is significant.
\end{abstract}

% Keywords

% Make Title Construct
\maketitle

% Section: Introduction
\section{Introduction}
\textit{Natural} or \textit{free} convection is the motion of a fluid around a body caused by density gradients in the presence of gravity. Common causes for density gradients could be heat or mass transfer. Many phenomena in geophysical flows and meteorology are governed by the principles of flow resulting from density stratification in the presence of gravity. Many engineering processes like vaporization, condensation and heat transfer in packed beds rely on natural convection. This has led to many detailed studies of buoyancy driven flows \cite{Turner73} and their hydrodynamic stability \cite{Chandrasekhar61}. Some exact solutions for flow in stratified fluids have also been derived \cite{Yih60,Drazin69}. Similarity solutions have been derived for buoyant plumes and thermals \cite{Batchelor54,Turner73} while a detailed theory for the phenomena of fluid entrainment by buoyant plumes was provided by Townsend in \cite{Townsend70}. 

Focusing on the case of heat transfer, temperature gradients can cause the distribution of the body force in the fluid to be non-uniform which in turn generates fluid motion. The density of the fluid in the low Mach number ($Ma$) limit is inversely related to the temperature via a multiplicative constant that is proportional to the thermodynamic pressure of the fluid. This dependence between density and temperature causes the fluid continuity, momentum, and energy equations to be coupled. The larger the magnitude of these temperature gradients, the stronger is the coupling between these equations. This coupling is expressed in terms of the Grashof number ($Gr$) which is the ratio  of the buoyancy force and the viscous force acting on a fluid. Closely related to $Gr$ is the Rayleigh number, $Ra = GrPr$, where $Pr$ is the Prandtl number. Coupling between the continuity, momentum, and energy equations renders their solution intractable to pen-and-paper analysis. However, some simplifying assumptions can be made when the governing parameters of the natural convection problem are constrained. 

Early solutions of the coupled equations either invoked the boundary layer approximation (suitable for large values of Reynolds number ($Re$) and/or $Gr$ justifying the thin boundary layer assumption while neglecting curvature effects) or linearized the governing equations paving the way for the method of matched asymptotic expansions \cite{Lagerstrom88}. Studies such as \cite{Merk53,Acrivos60nn,Potter80} focus on the limiting case of high $Gr$ using boundary layer assumptions. Classical analytical treatments of natural convection around spheres at small $Gr$ using matched asymptotic expansions may be found in \cite{Fendell68,Hieber69,Hossain70} while experimental studies of natural convection around a heated sphere at small $Gr$ have been conducted in \cite{Mathers57,Tsubouchi60,Yuge60} among others. These experiments were focused on measuring heat transfer and do not report measurements of drag on the sphere due to the fluid flow induced by natural convection. 

A widely used set of simplifying assumptions is the Boussinesq approximation \cite{Spiegel60} which is valid when $\Delta \rho/\rho \ll 1$, where $\rho$ denotes fluid density and $\Delta\rho$ denotes the change in density. A central theme of this paper is to study scenarios where the heat addition to a variable density fluid is not small and thus the Boussinesq approximation does not hold. Since the density is allowed to vary along with all other fluid properties, the flow does not remain incompressible but has a finite rate of dilatation. 

Spjut \cite{Spjut85} showed that the drag on a particle induced due to natural convection can be as great as the particle weight while \cite{Geoola81,Geoola82,Jia96} numerically calculated the drag due to natural convection on a heated sphere by solving the full Navier-Stokes equations in the regime where the Boussinesq approximation is valid. Dudek, et al. \cite{Dudek88} experimentally replicated the study of \cite{Geoola81,Geoola82} using an electrodynamic balance. 

The details of natural convective flows over surfaces and the characteristics of the resulting plume based on $Gr$ have been studied experimentally in \cite{Schenkels69,Jaluria75,Churchill02}. It is known from analytical solutions that in the absence of buoyancy forces ($Gr = 0$) and small $Re$ ($Re \ll O(1)$), the effect of heat transfer can be evaluated by computing the Nusselt number ($Nu$) which has a value of 2. As $Re$ increases, a $Re$ correction is required which was evaluated by \cite{CGW78}. In the limit of $Gr \rightarrow 0$, both perturbation and asymptotic expansion methods have failed \cite{CGW78} to yield solutions for $Nu$ with the generality of that obtained in \cite{Proudman57} for small Peclet numbers ($Pe$). Boundary layer approximations have been used for large $Gr$ \cite{Acrivos60,Stewart71,Chen78}. 

The setup of placing a heated sphere in a uniform flow has been termed \textit{forced} convection in literature. Many experiments and accompanying analytical studies have been carried out investigating both natural and forced convection on a variety of surfaces like spheres, flat plates, cylinders, sharp corners, and surface depressions when boundary layer approximations on these surfaces are valid. Forced convection in the absence of buoyancy with isothermal wall boundary conditions over a sphere and the resulting drag on the sphere has been studied in \cite{ganguli2019drag} using numerical simulations in the regime $10^{-3} \leq Re \leq 10$. 

The coupled problem of placing a heated sphere in a uniform flow in the presence of gravity is called \textit{mixed convection}. Scaling the terms in the momentum equation reveals that if $Gr/Re^2 \gg 1$, forced convection can be ignored; if $Gr/Re^2 \ll 1$, free convection can be ignored while if $Gr/Re^2 \approx 1$, the regime is that of combined forced and free convection. Depending on the domain of application and physical phenomena of interest, the effect of buoyancy can be scaled differently resulting in different non-dimensional numbers. 

In geophysics, the ratio of buoyancy and flow shear terms is called the Richardson number, $Ri = (g/\rho)((\partial \rho/\partial z)/(\partial u/\partial z)^2)$, where, $g$ is the acceleration due to gravity, $\rho$ is the ambient fluid density, $u$ is the fluid velocity, and $z$ is depth. In the design of chemical process reactors and fluidized beds, the ratio of buoyancy and viscous forces on a body is called the Archimedes number, $Ar = gL^3\rho\Delta\rho/\mu^2$, where, $\Delta \rho$ is the density difference in a fluid of ambient density $\rho$ owing to heat transfer due to a temperature difference $\Delta T$, $L$ is a characteristic length of the body, and $\mu$ is the fluid's dynamic viscosity. When the density difference, $\Delta \rho$, obeys $\Delta \rho/\rho = \beta \Delta T$, where $\beta$ is the volumetric thermal expansion coefficient, $Ar = Gr$ and $Ri = Gr/Re^2$. Thus, in thermal convection literature, $Ri$ has been interpreted as the ratio of gravity forces (corresponding to natural convection) and inertia forces (corresponding to forced convection).

In mixed convection, the direction of gravity and its relative orientation with respect to the direction of the freestream velocity plays an important role. When the buoyant motion is parallel to the direction of velocity, the flow is called \textit{aiding flow}. If the buoyant motion and forced motion are anti-parallel, the flow is called \textit{opposing flow}. In the limit of $Re \rightarrow 0$ and $Gr = O(Re^2)$, assuming the flow is incompressible and that Boussinesq approximations hold, the effect of aiding and opposing buoyancy on creeping flow has been obtained using the method of matched asymptotic expansions by \cite{Hieber69} and numerically by \cite{Woo71}. 

The unsteady counterpart of this problem with similar assumptions has been studied for moderate $Re$ and small $Gr$ such that $Gr/Re^2 \leq 40$. At higher $Re$, aiding flow delays the separation point further aft on the sphere while opposing flow moves the point forward \cite{Woo71}. Studies have documented the effects of aiding and opposing flow when $Re = O(100)$ and $O(100) < Gr < O(1000)$. Notable among these are \cite{Acrivos58,Yuge60,Pearson68} studying mixed convection at moderate $Re$ and $Gr$ on spheres, and \cite{Hatton70,Oosthuizen71} studying the corresponding behavior on cylinders. 

Motivated by recent interest in aerosol applications, flow features and dynamics due to mixed convection past a heated sphere have been studied numerically in the regime where Boussinesq assumptions hold for moderate $Re$ by \cite{Mograbi05a,Mograbi05b,Kotouc09a} while the transition to turbulence at higher $Re$ has been investigated in \cite{Kotouc09b}. The effect of $Pr$ and $Ri$ on mixed convection around a heated sphere at moderate $Re$ was studied by \cite{Nirmalkar13} for fluids with a power-law for viscosity variation assuming small density variations so that Boussinesq assumptions hold.  

It has been aptly stated in \cite{CGW78} that additional studies are needed to elucidate the intricate physical phenomena involved in mixed convection. In this paper, we investigate the effects of heat transfer and buoyancy on the drag of a spatially fixed heated spherical particle with isothermal wall boundary conditions in mixed convection at low $Re$ for a variable property fluid without making any approximations or assumptions on the resulting density variation or the magnitude of heat transfer from the sphere into the fluid. It is then instructive to ask: How is the drag on a heated sphere in a slow flow modified when gravity is present? Does the drag increase or decrease? Why? How does the orientation of gravity relative to the free-stream affect the drag? Fully resolved simulations are used to quantify the effects of heat transfer and buoyancy to answer the questions posed above. 

While preliminary simulations have been performed for cooled spheres, the results and analysis presented in this paper are only for heated spheres. The rest of the paper is organized as follows. Section \ref{governing_equations_notation} sets the stage by detailing the governing equations that are solved and the notation used in this paper. Section \ref{drag_modification} discusses the details of the observed drag modification of a heated particle in the presence of buoyancy using scaling arguments, parametric studies, the qualitative impact on flow features around the particle, and the model problem of a falling sphere under the influence of gravity. Section \ref{conclusions} summarizes the results along with concluding remarks.

% Section: Governing Equations and Formulation
\section{Governing Equations, Notation, and Numerical Solution}\label{governing_equations_notation}
\begin{figure}
	\centering
	\includegraphics[width=0.35\textwidth]{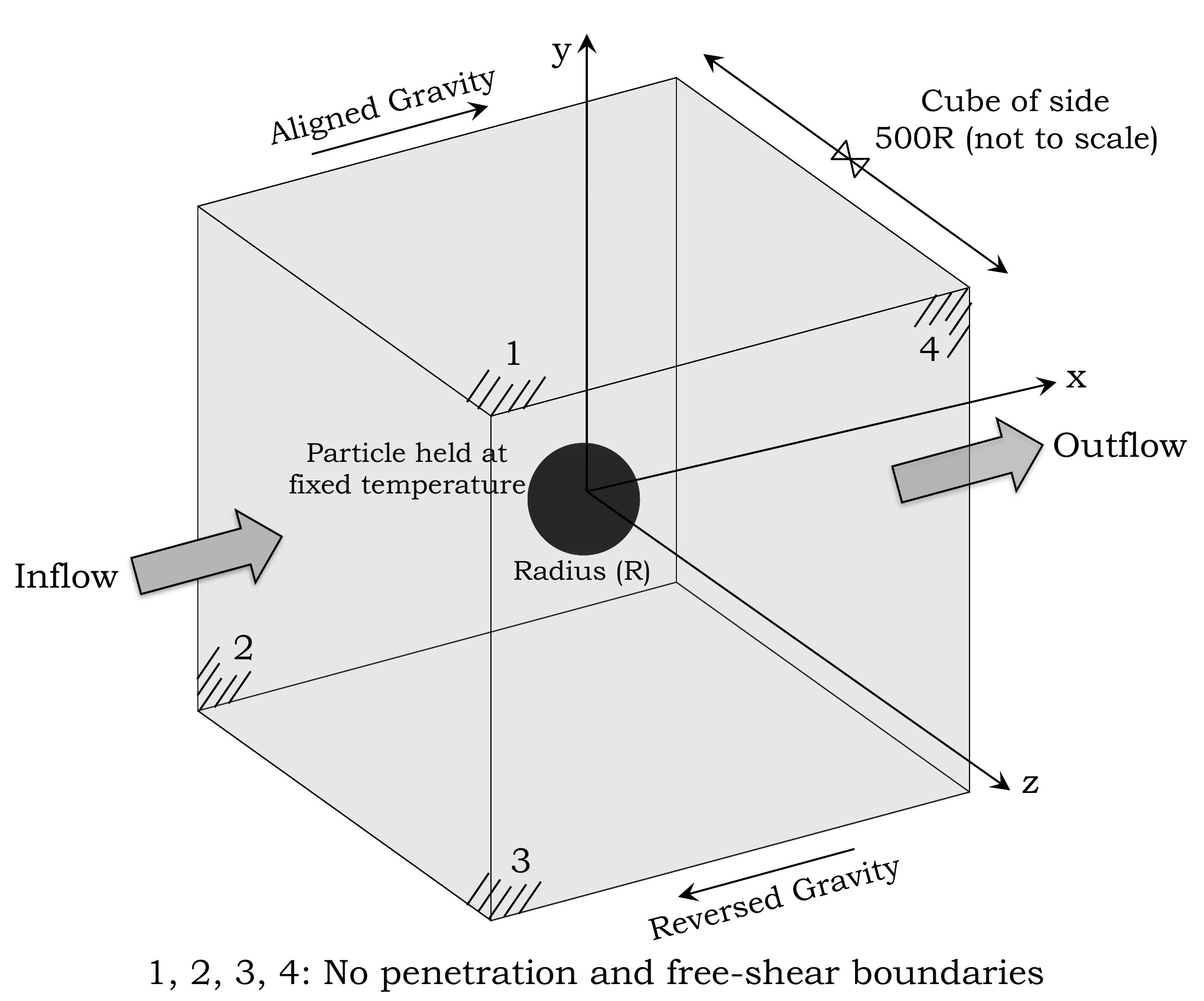}
	\caption{A schematic of the computational setup (not to scale). All contour plots in this paper are on the $xy$-plane in the reference frame shown with origin at the sphere's center.  \label{NumSetup}}
\end{figure}
A spatially fixed, heated spherical particle (radius $R$, diameter $D=2R$) is placed in a fluid with variable density ($\rho$), dynamic viscosity ($\mu$), and thermal conductivity ($\kappa$) which only vary with temperature ($T_g$) of the fluid. A uniform velocity, $U_\infty$, is prescribed far away from the fixed particle. Imposing a particle temperature, $T_p$, different from the ambient fluid temperature, $T_\infty$, at time $t=0$, compressibility effects are important and manifest as an acoustic front at small times \cite{ganguli2019low} while fluid motion occurs at much lower speed behind this front allowing for a low Mach number ($Ma$) limit ($Ma \rightarrow 0$) formulation of the Navier-Stokes equations \cite{Panton} which is used in this paper. 

The particle Biot number (ratio of heat conduction resistance to heat convection resistance) is assumed to be small ($Bi \ll 1$) implying that the particle is uniformly at the same temperature and cannot sustain any radial or angular (azimuthal or polar) gradients. Furthermore, conduction from the particle to the fluid and convection of the heat thereof is the only mode of heat transfer between the particle and the fluid. The fluid is assumed to be optically thin and in the absence of other particles, there is no scattered or incident radiation. An order of magnitude analysis \footnote{For air, $\kappa = 2 \times 10^{-2}W/mK$, $T_\infty = 300K$. For particle, $D=1\mu m$, emissivity, $\epsilon = 1$, Nusselt Number, $Nu=2$, $T_p=1200K$. $\sigma_{SB}$ is the Stefan-Boltzmann constant. Convective thermal power, $P_c = \kappa A Nu (T_p - T_\infty)/D$. Radiative thermal power, $P_r = \sigma_{SB} A (T_p^4 - T_{\infty}^4)$. Then $P_r/P_c = \mathcal{O}(10^{-3})$.} shows that heat transfer from the sphere due to radiation can be ignored compared to the heat transfer due to convection for the range of $T_p$ considered in this paper. The creeping flow limit ($Re$ is $\mathcal{O}(1)$ and smaller) is of special interest. 

Assuming the fluid as air ($Pr \approx 0.7$), the full 3D, variable density, low $Ma$ equations \cite{Panton} delineated below are solved for the fluid using the approach in \cite{Ham07} without making any assumption on the amount of heat addition from the sphere into the fluid or the relative magnitude of the body force with respect to the forces solely due to fluid motion (characterized by the Froude number, $Fr$).
\begin{subequations}
    \begin{gather}
        \partial_t \rho + \partial_{x_j} (\rho u_j)  = 0 \label{continuity} \\
        \partial_t (\rho u_i) + \partial_{x_j} (\rho u_i u_j) = -\partial_{x_i} p + \partial_{x_j}\tau_{ij} + \rho g_i \label{momentum} \\
        \tau_{ij} = \mu\left(\partial_{x_j} u_i + \partial_{x_i} u_j - 2/3\delta_{ij} \partial_{x_k} u_k \right) \label{stress} \\
        \partial_t (\rho C_v T_g) + \partial_{x_j} (\rho C_p T_g u_j) = \partial_{x_j}\left(\kappa \partial_{x_j} T_g\right) \label{energy} \\
        P = P_0 + p = \rho R T_g \label{state}
    \end{gather}
\end{subequations}
In the above equations, $u_i$ is the $i^{\text{th}}$ component of the fluid velocity vector $\vec{u}$ (the $x$, $y$, $z$ components are also denoted by $u$, $v$, and $w$, respectively), $p$ is the sum of the hydrostatic and hydrodynamic pressure, $P_0$ is the thermodynamic pressure, $R$ is the ideal gas constant for air, $g_i$ is the component of gravity in the $i^{th}$ direction, $C_v$ is the isochoric specific heat capacity, and $C_p$ is the isobaric specific heat capacity. The total pressure is $P$ where, in the low $Ma$ limit, $p \ll P_0$. $C_p$ and $C_v$ are assumed to be constants. 

In the limit $Ma \rightarrow 0$, the second term of the energy equation includes the pressure work term in addition to the internal energy which evaluates to the enthalpy $C_pT_g$ \cite{Panton}. Heating due to viscous dissipation is negligible \cite{Kundu51}. A power law \cite{Vincenti65} is used to model the dynamic viscosity as a function of temperature given by $\mu = \mu_0\left(T_g/T_0\right)^n$, where $\mu$ is the viscosity at temperature $T_g$ and $\mu_0$ is the reference viscosity at a reference temperature $T_0$. For air, $\mu_0 = 1.716$ x $10^{-5}$ kg/m-s, $T_0 = 273$ K, and $n = 2/3$. The variation of thermal conductivity is given by $\kappa(T_g) = \mu(T_g)C_p/Pr$. 

It must be noted that there are other widely used correlations for $\mu$ and $\kappa$ of air in the range of temperatures of interest in this paper which may be more accurate at higher temperatures and therefore induce small discrepancies in the drag values reported in this paper at these temperatures. The specific heats and $Pr$ are also functions of temperature, albeit the variation of these quantities in the range of temperatures of interest in this paper are below $10\%$ and will only result in small discrepancies in the drag values reported in this paper. However, these small discrepancies do not affect the analysis presented in this paper. 

Figure \ref{NumSetup} shows a schematic of the numerical setup which consists of a sphere of radius $R$ held spatially fixed at a fixed temperature inside a cubic box large enough such that $99\%$ of the free-stream velocity is recovered at the outlet for the $Re$ range of interest in this paper. ($\hat{i}, \hat{j}, \hat{k}$) denote the unit vectors along the ($x,y,z$) axes in the reference frame shown in the figure, respectively. All contour plots in this paper are on the $xy$-plane in the reference frame shown with origin at the sphere's center. All descriptions of the orientation of vectors (for example, forces, gravity, velocity, etc.) are in the reference frame shown in figure \ref{NumSetup}. Except for the presence of the body force term in the momentum equation, the computational setup is identical to \cite{ganguli2019drag}. The reader is referred to \cite{ganguli2018computational} for more details on the choice of the size of the domain, construction of the computational mesh, and verification and validation of the numerical code.

\begin{figure}
	% \addtolength{\subfigcapskip}{-0.1in}
    \centering
	\subfigure[$Re$ = 0.01]{\includegraphics[width=0.155\textwidth]{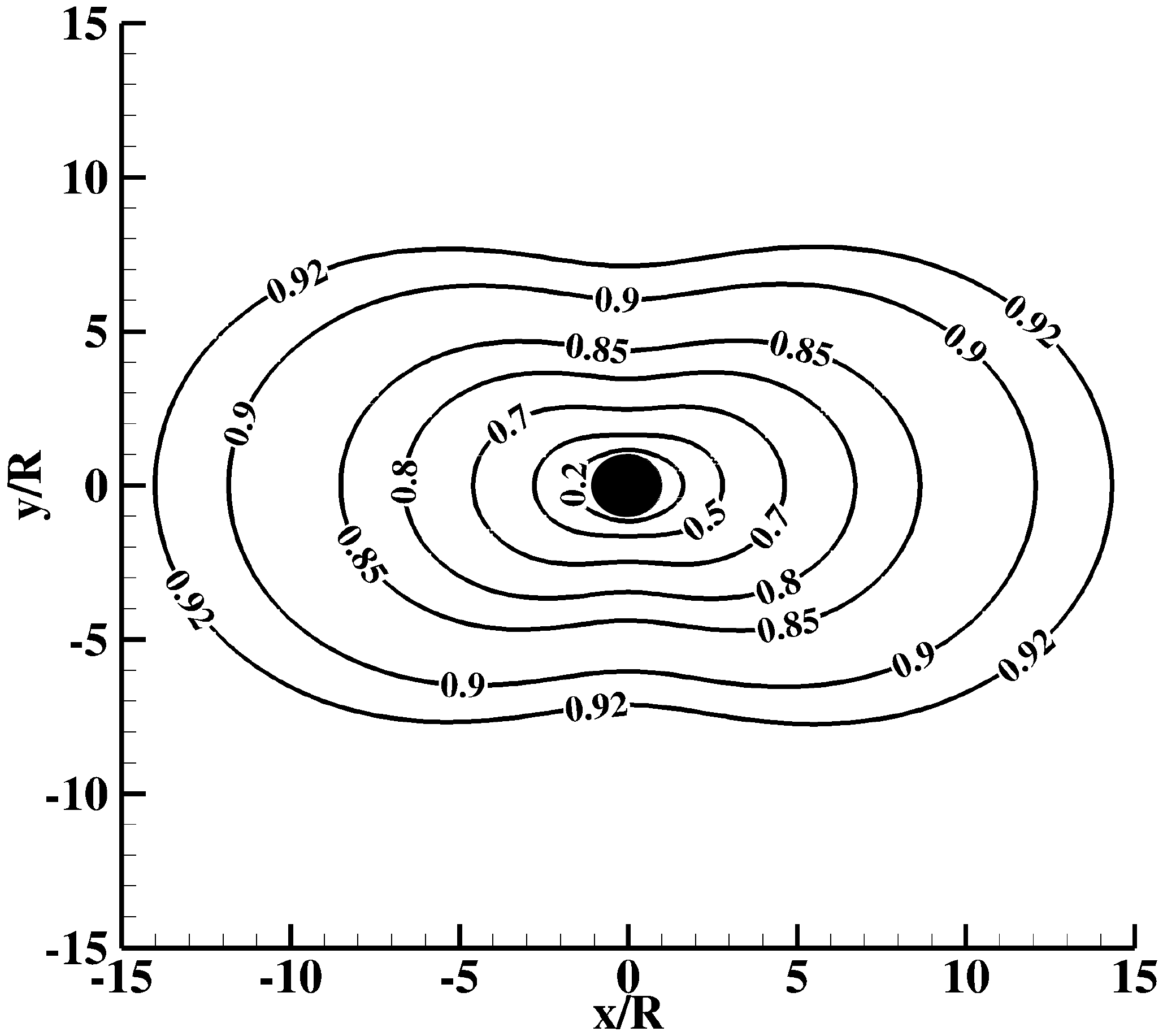}}
	\subfigure[$Re$ = 0.1]{\includegraphics[width=0.155\textwidth]{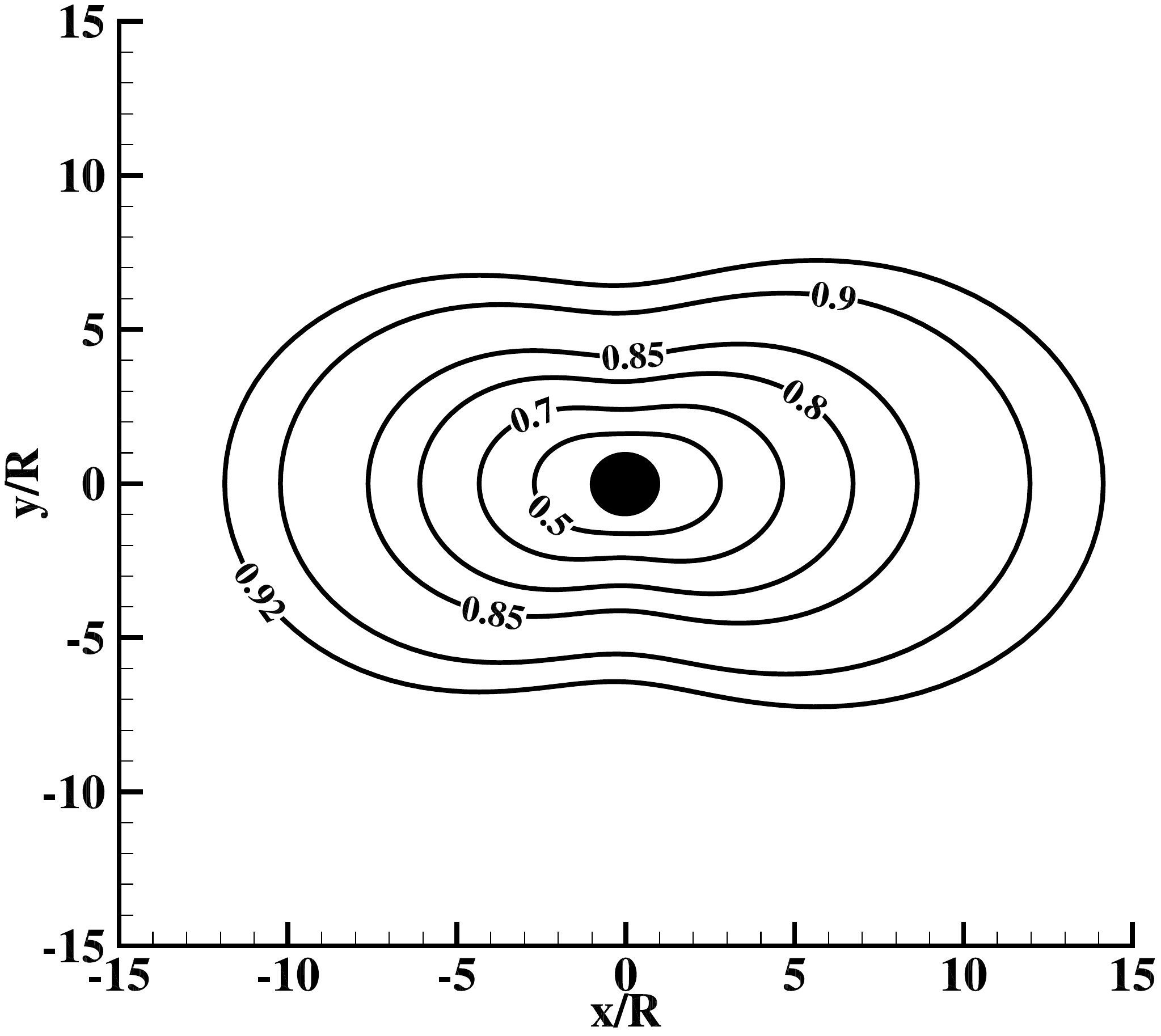}}
	\subfigure[$Re$ = 1]{\includegraphics[width=0.155\textwidth]{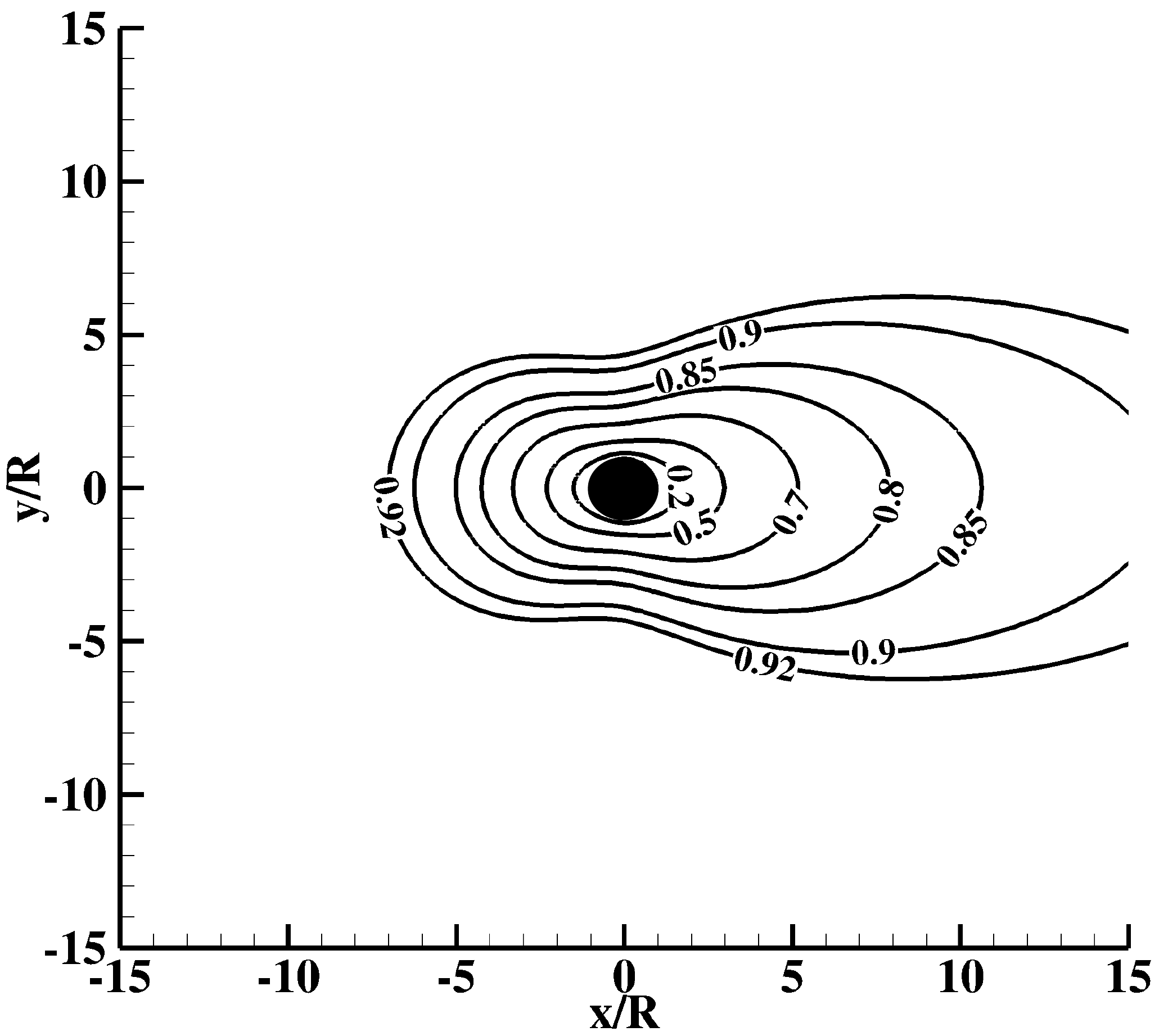}}
	\caption{Contours of normalized $x$-component of the velocity, $u/U_{\infty}$, for an unheated ($\lambda = 0$) sphere for $Re = 0.01, 0.1$, and $1$ in the absence of gravity ($Fr = \infty$).}
	\label{Isothermal}
\end{figure}

Our study aims to quantify the effect of heat transfer in the presence of gravity on the drag of a heated sphere in the low $Re$, low $Bi$, and low $Ma$ regime. The governing parameters of the problem are the inflow Reynolds number based on $D$ ($Re$), the Boussinesq parameter ($\lambda$), and the Froude number ($Fr$). The inflow Reynolds number is defined as $Re = \rho_{\infty} U_{\infty} D/\mu_{\infty}$. The Boussinesq parameter \cite{Lagerstrom64,Panton} is defined as the difference between the particle temperature, $T_p$, and the ambient temperature of the gas, $T_\infty$, normalized by $T_\infty$ so that $\lambda = (T_p-T_\infty)/T_\infty$. The Froude number is defined as the ratio between inertial and buoyancy forces evaluated in the presence of heat transfer so that $Fr = U_\infty/\sqrt{\lambda g D}$. The subscript $\infty$ denotes the values of the parameters in the far-field. 

Note that $\lambda = 0$ corresponds to isothermal conditions (which, in the absence of buoyancy, is Stokes' flow) and $\lambda \ll  1$ corresponds to the Boussinesq regime where small temperature changes cause small density changes. In the absence of buoyancy, $\lambda = 0.1$, $\max(\Delta \rho/\rho) \sim 10\%$ while at $\lambda = 0.2$, $\max(\Delta \rho/\rho) \sim 18\%$ meaning that non-Boussinesq effects become dominant when $\lambda > 0.1$. 

The parameter specification and numerical formulation of the code are non-dimensional thereby yielding generalizable results. The pressure drag ($\vec{F}_{P}$) and viscous drag ($\vec{F}_{V}$) is evaluated on the surface of the sphere, $\partial S$, as $\vec{F}_{P} = \oint_{\partial S} (p-p_{\infty})\bar{\bar I} \cdot \vec{n}\,\,\,\,dS$ and $\vec{F}_{V} = \oint_{\partial S} \bar{\bar \tau} \cdot \vec{n}\,\,\,\,dS$, respectively, where $\vec{n}$ is the surface normal vector. In literature, reporting the non-dimensional drag as the coefficient of drag using the inertial scaling is most popular. We will adopt this convention for reporting the computed drag values. An alternative approach is to use the coefficient of drag using the viscous scaling. Both are however equivalent by a multiplicative constant of $Re/4$. Letting $S_I = (1/2)\rho_{\infty} U_{\infty}^2 \frac{\pi}{4} D^2$, the inertially scaled coefficient of drag may be computed as $C_D = (|\vec{F}_{P} + \vec{F}_{V}|)/S_I$. 

In the sections that follow, comparisons are made between the drag of an unheated sphere with that of a heated sphere. For the heated sphere, the drag is calculated for forced, natural, and mixed convection setups. $C_D^{0} = C_D^0(Re)$ denotes the drag on an unheated sphere and is a function only of $Re$. $C_D^F = C_D^F(Re,\lambda)$ denotes the drag of a heated sphere in forced convection, $C_D^N = C_D^N(\lambda, Fr)$ denotes the drag in natural convection, while $C_D^M = C_D^M = C_D^M(Re, \lambda, Fr)$ denotes the drag in mixed convection. While values of $C_D^0$ and $C_D^F$ are directly available from tables 2 and 3 in \cite{ganguli2019drag}, correlations can also be used. Given the range of $Re$ and $\lambda$ of interest in this paper, we use the Clift-Grace-Weber correlation \cite{CGW78} for $C_D^0$ and the correlation proposed in \cite{ganguli2019drag} for $C_D^F$ which are $C_D^0 = 24(1+0.1315Re^{0.82-0.05\log_{10}Re})/Re$ and $C_D^F = C_D^0 + 10.7672\lambda^{0.9673}Re^{-0.9529}$, respectively.

Since the effects of the orientation of gravity is also studied, the drag coefficient is associated with a sign denoting directionality. Drag coefficients associated with forces along the $+x$ or $+y$ axes are denoted with a $+$ sign while those associated with forces along the $-x$ or $-y$ axes are denoted with a $-$ sign. In section \ref{perpendicular_gravity_section}, $C_D^x$ and $C_D^y$ denote the components of the computed total drag in mixed convection along the $x$ and $y$ axes, respectively. 

To aid the explanation of and to help contrast the effects of buoyancy on the features of the flow around the heated sphere, which are shown via contour plots presented later in this paper, contours of the normalized $x$-component of velocity ($u/U_\infty$) for Reynolds numbers $Re = 0.01$, $Re=0.1$, and $Re = 1$ are shown in figure \ref{Isothermal} when the sphere is not heated and gravity is absent. The corresponding contours of pressure and vorticity may be found in \cite{ganguli2018computational}.

% Section: Drag Modification of a Heated Sphere in the Presence of Gravity
\section{Drag Modification of a Heated Sphere in the Presence of Gravity}\label{drag_modification}
When a sphere immersed in a variable density fluid is heated, the fluid in the near vicinity of the sphere is warmer than the surrounding fluid and therefore has a lower density. In the presence of gravity, this warm fluid has the tendency to migrate in the direction opposite to gravity as the colder heavier fluid occupies its position. This migration of the fluid imparts a viscous pulling force on the sphere in addition to a force due to the additional pressure difference caused by the fluid movement. The viscous pulling on the sphere reduces the inertial drag on the particle from the convective flow when the direction of gravity is aligned to the direction of the background flow while it enhances drag on the sphere when the direction of gravity is reversed from the direction of the background flow. Depending on the reference pressure chosen for $x = 0$, presence of buoyancy also adds a hydrostatic component to the pressure at every point in the domain. There is also a net pressure force on the body which is the Archimedian buoyancy force. 

The contribution of these forces depends on the $Re$, $\lambda$, and $Fr$ of the flow and the gradients in the variable properties of the fluid. The variable properties are assumed to vary only with temperature in this paper and thus they vary only as a function of $\lambda$. These forces need to be incorporated into the total drag calculation while also assessing the partial contributions from each of the components. 

A parametric study using fully resolved simulations of the heated particle is carried out to obtain the modified component-wise drag when the direction and magnitude of gravity is varied. The direction of the gravity vector may be chosen by specifying its components along the three axes in the reference frame shown in figure \ref{NumSetup}. For all computations presented in this paper, the background flow velocity is $U_\infty\hat{i}$, where $U_\infty$ is determined by $Re$. 

In this paper, we focus our attention on three canonical cases: (i) \textit{Aligned Gravity}: The direction of gravity is aligned with the direction of the far-field uniform flow ($\vec{g} = |\vec{g}|\hat{i}$) and thus the flow induced by gravity opposes the background uniform flow, (ii) \textit{Reversed Gravity}: The direction of gravity is anti-parallel to the far-field uniform flow ($\vec{g} = -|\vec{g}|\hat{i}$) and thus the flow induced by gravity aids the background uniform flow, and, (iii) \textit{Perpendicular Gravity}: The direction of gravity is perpendicular to the direction of the far-field uniform flow ($\vec{g} = |\vec{g}|\hat{j}$). In the range of $Re$, $\lambda$, and $Fr$ investigated in this paper, the direction of gravity influences the sign of the buoyancy correction term. 

Although both aligned and reversed gravity setups are simulated, the aligned gravity setup is presented in more detail compared to the reversed gravity setup for brevity since the observations made in both setups are similar. Fully resolved simulations solve the full Navier-Stokes equations (\ref{continuity}), (\ref{momentum}), (\ref{energy}), and (\ref{state}) in three dimensions with the body force term added.

\subsection{Scaling the viscous pull opposite to gravity}\label{scaling_viscous_pull}
\begin{figure}
	\centering
	\subfigure[$Fr = 10$]{\includegraphics[width=0.155\textwidth]{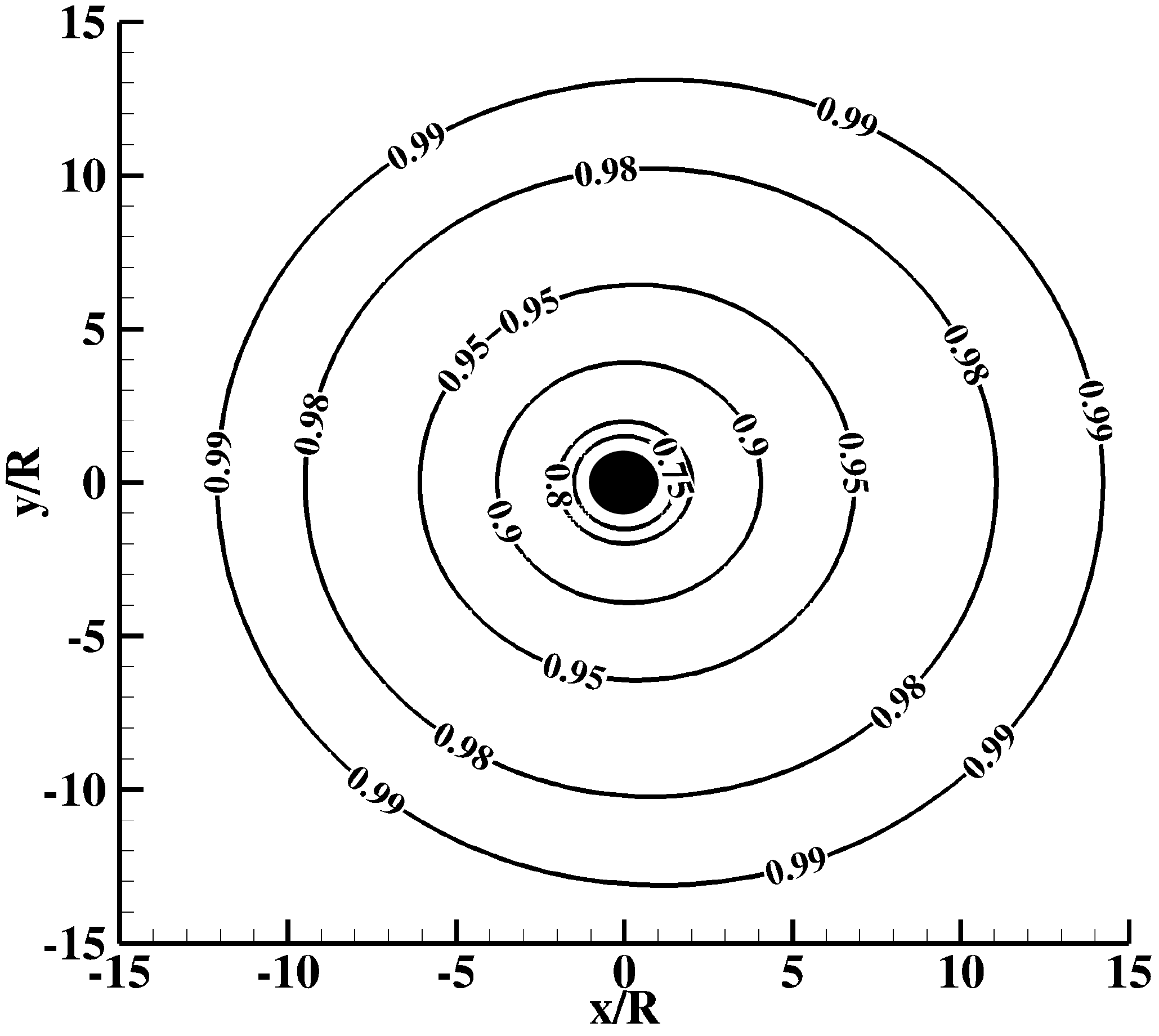}}
	\subfigure[$Fr = 1$]{\includegraphics[width=0.155\textwidth]{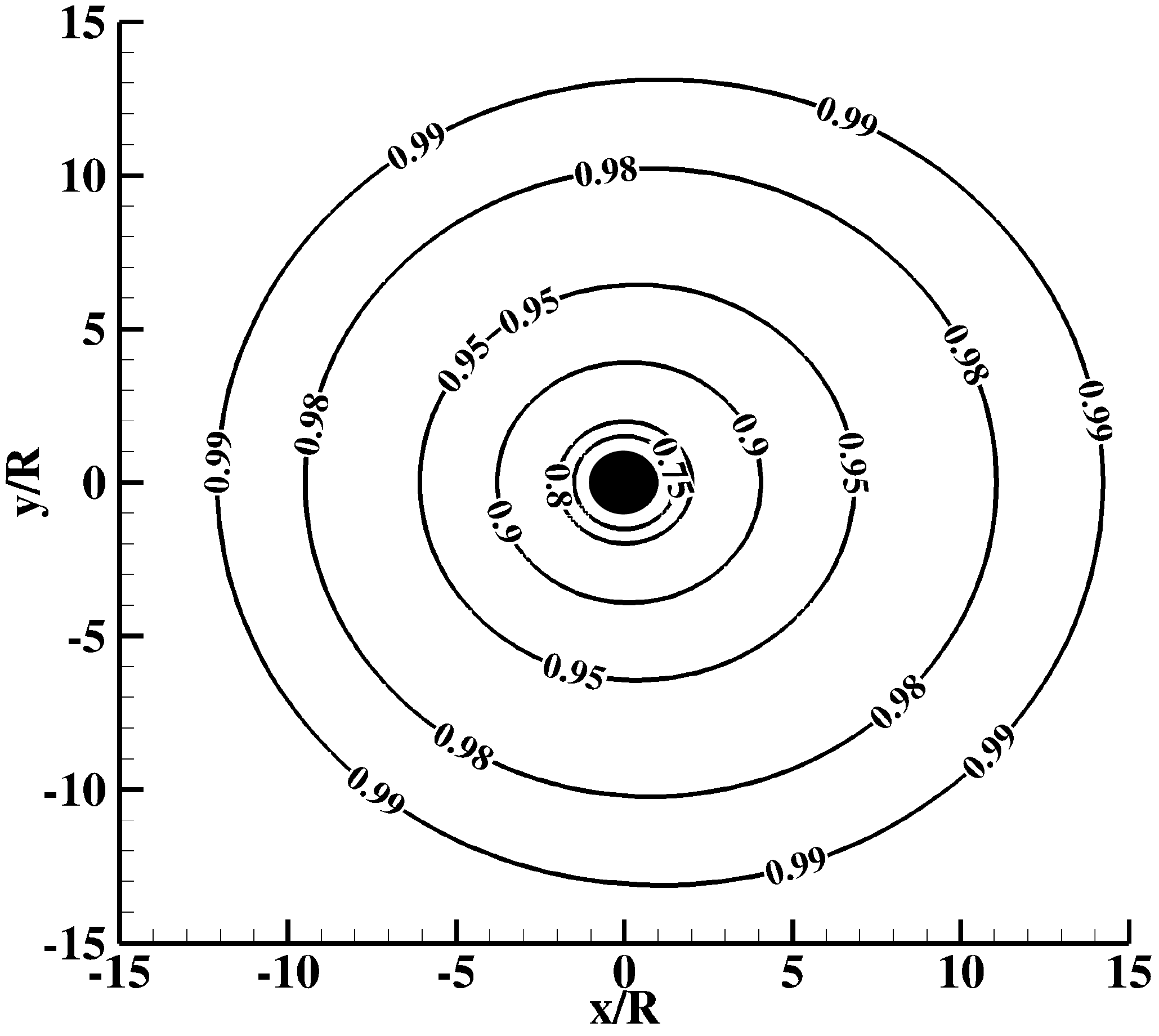}}
    \subfigure[$Fr = 0.1$]{\includegraphics[width=0.155\textwidth]{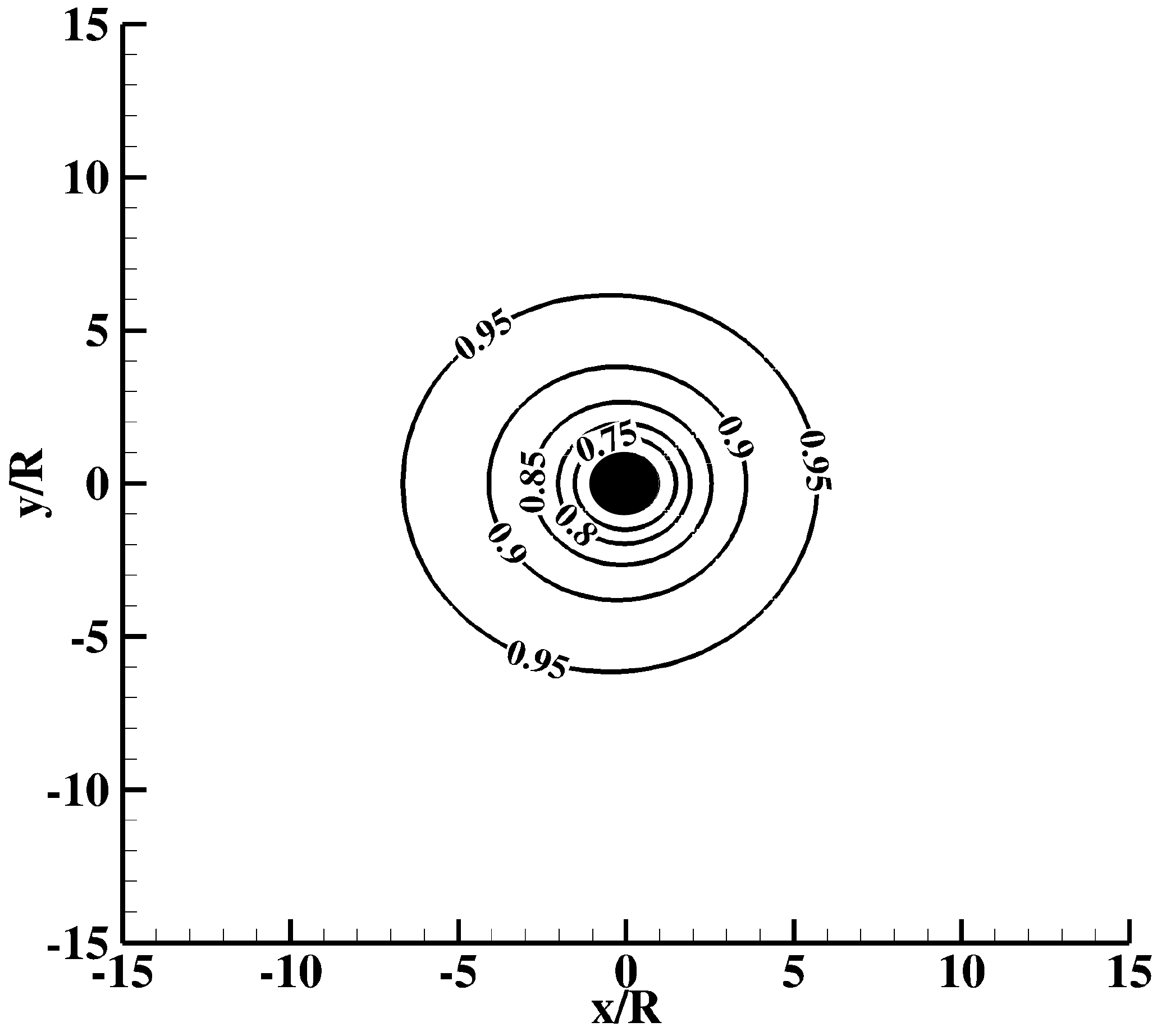}}
	\caption{Contours of normalized density, $\rho/\rho_{\infty}$, over a 3 decade variation in $Fr$ (equivalently $Re_{BI}$) while $Re$ and $\lambda$ are fixed at $Re=0.1$, $\lambda=0.5$ in the aligned gravity setup.\label{dens_g_contours}}
\end{figure}

Section 3.1 in \cite{ganguli2019drag} demonstrates that variable density effects are important in a fluid when $\beta_\infty \lambda T_\infty Re/\gamma_\infty  \gg Ma^2$, where $\beta_\infty$ is the fluid's bulk expansion coefficient evaluated at the far-field temperature. Specifically, it is shown that variable density effects are important in air. An exhaustive set of simulations, detailed in \cite{ganguli2018computational}, were carried out to assess the steady state density changes that result around a heated sphere placed in a uniform flow in the presence of gravity as $Re$, $\lambda$, $Fr$, and the direction of gravity are varied. Figure \ref{dens_g_contours} shows contour plots of the appropriately normalized density for three chosen cases while contour plots for other simulated cases are documented in \cite{ganguli2018computational}. 

It is observed that when $Re < 0.1$ and $0 \leq \lambda \leq 2$, the density field remains close to spherical (eccentricity of the contours is below 0.07) despite a three decade variation in $Fr$ from 0.1 to 10. The temperature field mimics this behavior since in the limit $Ma\rightarrow0$, the total pressure is almost constant and dominated by the thermodynamic pressure leading to temperature and density varying inversely following equation (\ref{state}). The small eccentricity is a manifestation of the Oseen correction terms (small convective corrections) which are important when $r\sim\mathcal{O}(R/Re)$ ($r$ being the radial coordinate with respect to the center of the sphere) in the absence of which, one would be solving a simple heat equation with spherically symmetric boundary conditions yielding exactly spherical contours. 

This observation suggests that the convective terms in the momentum equation (\ref{momentum}) takes the form of linearized Oseen-like correction terms so that in the low $Re$, low $Ma$ limit, the steady state momentum equation can be simplified to
\begin{equation}\label{momimp}
    \rho U_\infty \vec{i} \cdot \nabla \vec{u} + \nabla p = \nabla \cdot \bar{\bar{\tau}} + \rho \vec{g}
\end{equation}
where, $\bar{\bar{\tau}} = \mu\left(\nabla \vec{u} + (\nabla \vec{u})^T - 2/3\left(\nabla \cdot \vec{u}\right)\bar{\bar{I}}\right)$ and $\bar{\bar{I}}$ is the identity tensor. A scaling analysis of each term in equation (\ref{momimp}) can help understand the conditions under which these contributions balance and conditions under which a particular contribution dominates. At small $Re$, the Oseen correction terms are small compared to the other terms in equation (\ref{momimp}). Assuming the orientation of gravity is either aligned or reversed compared to the background uniform flow, the buoyancy term, $B = \rho \vec{g}$, induces migration of the warm and lighter fluid in the near vicinity of the sphere in the direction opposite to $\vec{g}$, thereby imparting a viscous pulling force on the sphere. 

Define the characteristic fluid velocities due to this induced motion as the \textit{buoyancy induced viscous velocity}, $U_{BV}$, and the \textit{buoyancy induced inertial velocity}, $U_{BI}$, which are obtained by scaling $B$ with the viscous and pressure contributions in equation (\ref{momimp}), respectively. Thus, $\lambda g \sim U_{BV} \nu_{\infty}/D^2 \implies U_{BV} \sim \lambda g D^2/\nu_{\infty}$ and $\lambda g \sim U_{BI}^2/D \implies U_{BI} \sim \sqrt{\lambda g D}$. Based on these velocity scales, define the \textit{buoyancy induced viscous Reynolds number}, $Re_{BV}$, and the \textit{buoyancy induced inertial Reynolds number}, $Re_{BI}$, as $Re_{BV} = U_{BV}D/\nu_{\infty} = \lambda g D^3/\nu_{\infty}^2 = \left(Re/Fr\right)^2$ and $Re_{BI} = U_{BI}D/\nu_{\infty} = \sqrt{\lambda g D^3}/\nu_{\infty} = Re/Fr$, respectively. This allows for conveniently relating all governing parameters $Re$, $Fr$, and $\lambda$ of the mixed convection problem via either of $Re_{BI}$ or $Re_{BV}$ as $Re_{BV} = \left(Re/Fr\right)^2 = Re_{BI}^2$. $Re_{BI}$ allows us to compress the large parametric space of $Re$, $Fr$ and $\lambda$ into a single parameter. 

Note that the form of $Re_{BV}$ is very similar to that of the Grashof number, $Gr$, which is the dimensionless number denoting the ratio of buoyancy and viscous forces and is used widely in problems involving natural convection. The Grashof number is $Gr = g\beta(T_p - T_{\infty})D^3/\nu_{\infty}^2$ where $\beta = \beta(T) = -1/\rho\left(\partial\rho/\partial T\right)_p$ is the coefficient of thermal expansion of the fluid and is a strong function of temperature. For ideal gases, $\beta \approx 1/T_g$ and thus $Re_{BV} \approx Gr$. However, this is not true for other fluids and the functional relationship of $\beta$ on $T$ could be more complex. For this reason, we will use $Re_{BV}$ (or equivalently $Re_{BI}$) instead of $Gr$ in our analysis.

\subsection{Mixed convection as linear superposition of forced and natural convection at low $Re$}
\begin{figure*}
	\centering
	% \subfigure[$Re, \lambda, Re_{BI} = 0.1$]{\includegraphics[width=0.22\textwidth]{Figures_PaperIII/Figure_4a.pdf}}
	% \subfigure[$Re, \lambda = 0.1$ \& $Re_{BI} = 1$]{\includegraphics[width=0.22\textwidth]{Figures_PaperIII/Figure_4b.pdf}}
    % \subfigure[$Re = 1$ \& $\lambda, Re_{BI} = 0.1$]{\includegraphics[width=0.22\textwidth]{Figures_PaperIII/Figure_4c.pdf}}
    % \subfigure[$Re, Re_{BI} = 0.1$ \& $\lambda = 0.5$]{\includegraphics[width=0.22\textwidth]{Figures_PaperIII/Figure_4d.pdf}}
    
    % \subfigure[$Re, \lambda, Re_{BI} = 0.1$]{\includegraphics[width=0.22\textwidth]{Figures_PaperIII/Figure_4e.pdf}}
	% \subfigure[$Re, \lambda = 0.1$ \& $Re_{BI} = 1$]{\includegraphics[width=0.22\textwidth]{Figures_PaperIII/Figure_4f.pdf}}
    % \subfigure[$Re = 1$ \& $\lambda, Re_{BI} = 0.1$]{\includegraphics[width=0.22\textwidth]{Figures_PaperIII/Figure_4g.pdf}}
    % \subfigure[$Re, Re_{BI} = 0.1$ \& $\lambda = 0.5$]{\includegraphics[width=0.22\textwidth]{Figures_PaperIII/Figure_4h.pdf}}
    
    \includegraphics[width=0.76\textwidth]{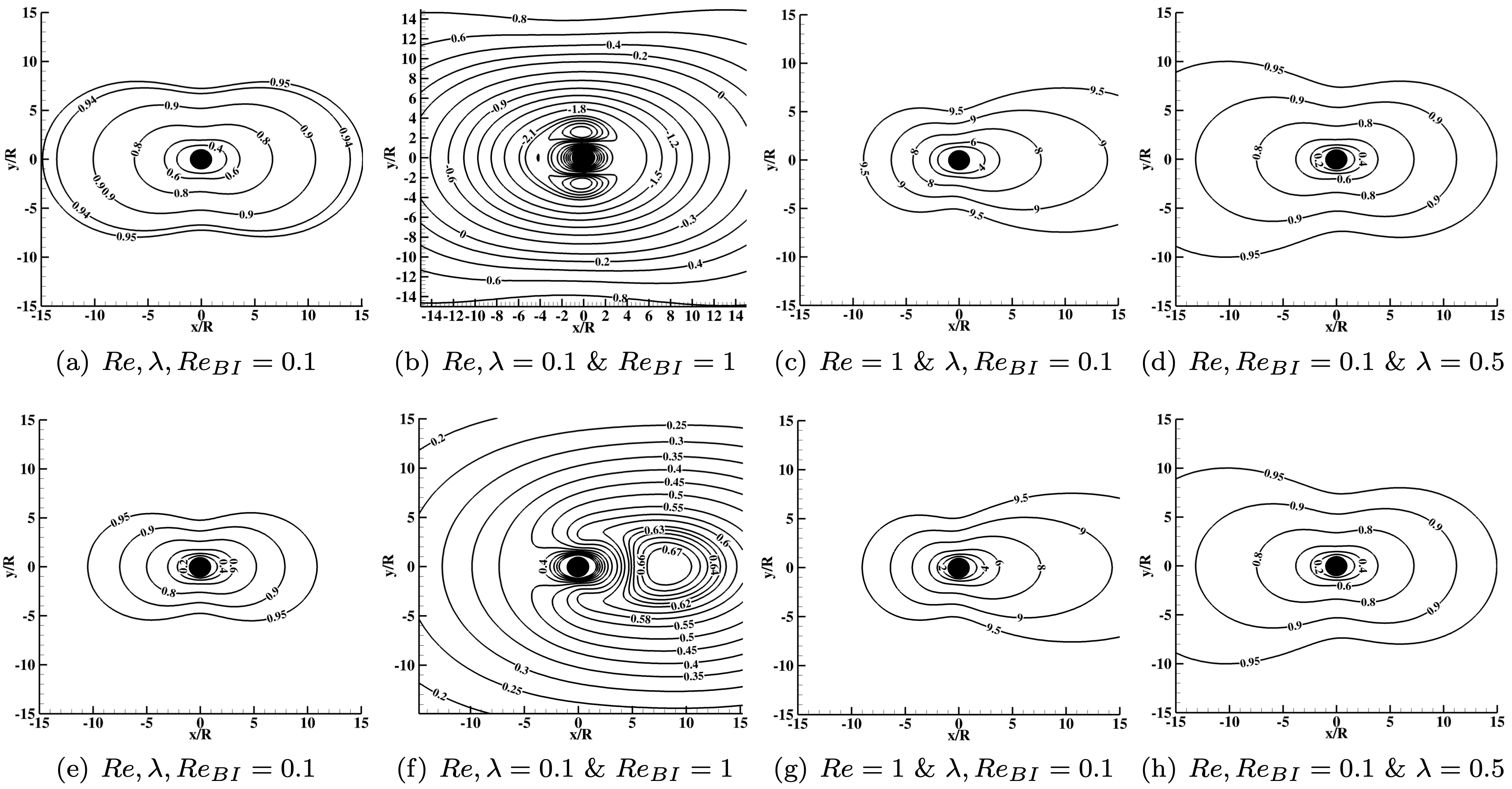}
    
	\caption{Contours of normalized $x$-component of velocity, $u/\sqrt{\lambda g D}$. The top and bottom rows shows the contours in the aligned and reversed gravity setups, respectively. Figures (a) and (e) are contours for a base case when $Re, \lambda, Re_{BI} = 0.1$. Figures (b),(f) demonstrate the effect of buoyancy by increasing $Re_{BI}$ to 1, figures (c),(g) demonstrate convective effects by increasing $Re$ to 1, and figures (d),(h) demonstrate heat transfer effects by increasing $\lambda$ to 0.5.\label{u_g_contours}}
\end{figure*}
\begin{figure}
    \centering
    \includegraphics[width=0.40\textwidth]{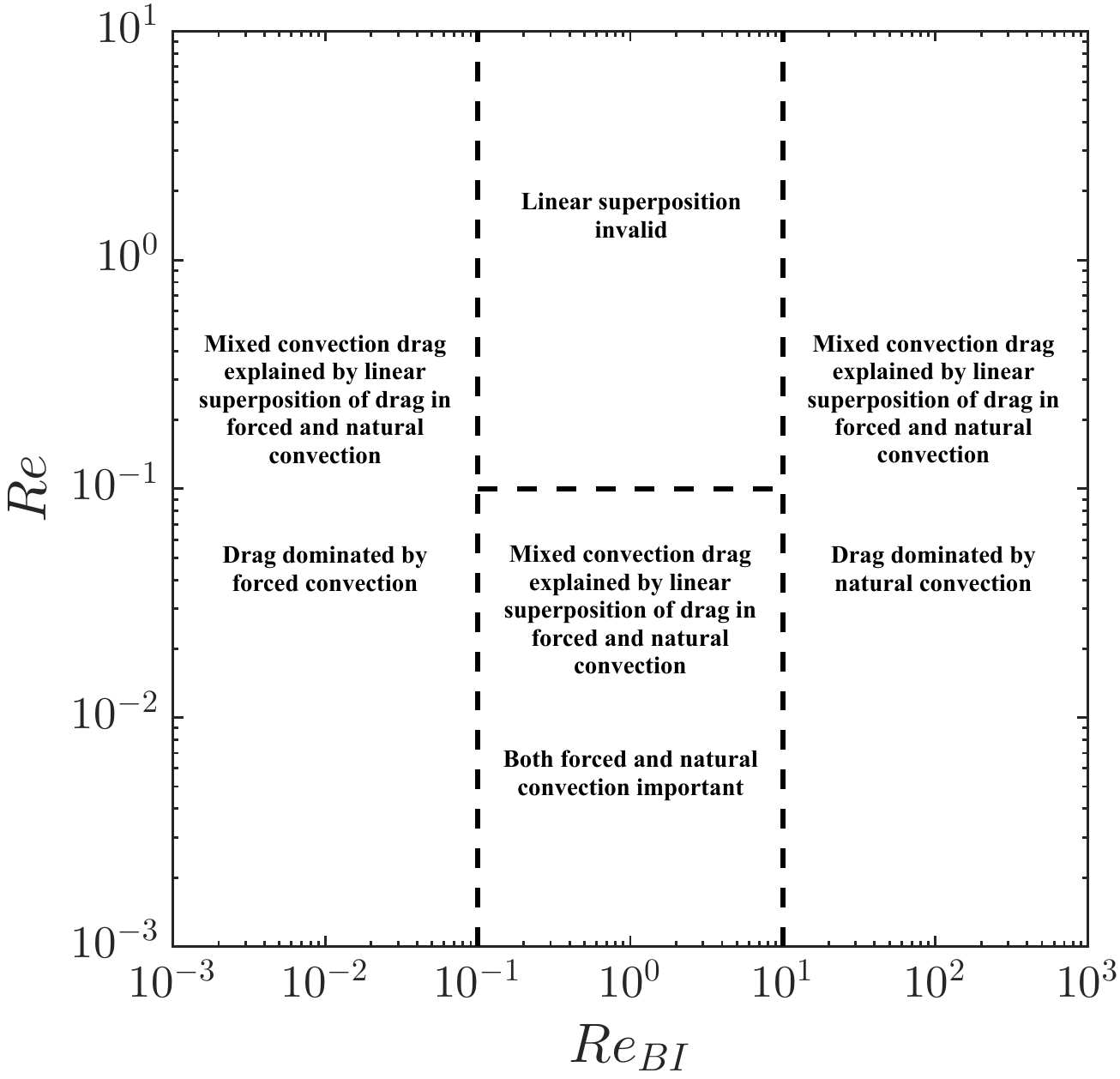}
    \caption{Summary of linear superposition analysis of drag on a sphere in mixed convection in $Re$-$Re_{BI}$ parameter space.}
    \label{summary}
\end{figure}

A parametric study in the range $10^{-3} \leq Re \leq 10$, $0 \leq \lambda \leq 3$, and $0.1 \leq Fr \leq 10$ of the governing parameters using fully resolved simulations of the heated particle is carried out to obtain the detailed breakdown of component-wise drag in mixed convection. The corresponding cases of natural convection are also simulated while the corresponding case of forced convection is available from analysis carried out in \cite{ganguli2019drag}. The results from these simulations are tabulated in tables \ref{draggravity}, \ref{draggravity_unsuccessful}, and \ref{draggravity_reversed} in section \ref{tables_of_drag} of the appendix. Tables \ref{draggravity} and \ref{draggravity_unsuccessful} summarize the results for aligned gravity cases while table \ref{draggravity_reversed} summarizes the results for reversed gravity. In addition to the drag in mixed convection, the tables also show the corresponding values for the drag on the sphere in forced (same $Re$ and $\lambda$ as in mixed convection) and natural convection (same $\lambda$ and $Fr$ as in mixed convection). All drag coefficient values use the inertial scaling. The qualitative observations from tables \ref{draggravity}, \ref{draggravity_unsuccessful}, and \ref{draggravity_reversed} and linear superposition analysis presented in this section has been summarized in figure \ref{summary}.

It is observed that in the regime $\{\{Re_{BI} < 0.1\} \cup \{Re_{BI} > 10\} \cup \{\{0.1 < Re_{BI} < 10\} \cap \{Re < 0.1\}\}\}$, the drag on a heated sphere in mixed convection at a given $Re$, $\lambda$, and $Fr$ can be analyzed and understood as the superposition of the drag on a heated sphere in uniform flow (forced convection) with the same $Re$ and $\lambda$ as in mixed convection, and the drag on a heated sphere in natural convection in the presence of gravity at the same $\lambda$ and $Fr$ as in mixed convection. Note that the drag reported in natural convection, $C_D^N$ is the sum of the hydrostatic component and the hydrodynamic (inertial and viscous) components. From the scaling analysis in section \ref{scaling_viscous_pull} on the terms in equation (\ref{momimp}), it is seen that the Oseen correction term scales as $Re^2$, the pressure and viscous terms scale as $Re$ while the buoyancy term scales as $Re_{BV} = Re_{BI}^2$. 

Based on this scaling, we expect that when $Re < O(1)$, the Oseen correction term is small and the pressure and viscous contributions are comparable. When $Re \gg Re_{BI}^2$, the presence of gravity does not influence the total drag and thus forced convection serves as a good approximation to mixed convection. On the other hand, when $Re \ll Re_{BI}^2$, the contribution from gravity dominates the total drag and thus natural convection serves as a good approximation to mixed convection. Our fully resolved numerical simulations corroborate these expectations as can be seen from tables \ref{draggravity} and \ref{draggravity_reversed}. The regime where $0.1 < Re_{BI} < 10$ is a regime where both forced and natural convection contributions are important and the numerical simulations help shed light on how they measure up to explain the drag in mixed convection. From the tables in section \ref{tables_of_drag} of the appendix, it is seen that the drag on a heated sphere in mixed convection can be described within $\pm10$\% error by delineating three distinct regimes determined by the value of $Re_{BI}$. 

The first regime is when $Re_{BI} < 0.1$, which is the lower extremal regime. In this regime, $Fr$ is large compared to $Re$. This means that convective effects dominate and overpower buoyancy effects. Flow of the fluid around the particle is strongly modified in the presence of heat transfer from the particle as demonstrated in \cite{ganguli2019drag}. The drag experienced in forced convection accounts for more than 95\% of the drag in mixed convection implying that drag modification is dominated by heat transfer effects compared to buoyancy effects. 

The second regime is when $Re_{BI} > 10$, which is the higher extremal regime. In this regime, $Fr$ is small compared to $Re$ and thus the buoyancy effects overpower convective or heat transfer effects. The drag experienced in natural convection explains more than 95\% of the total drag on the particle in mixed convection implying that drag modification is dominated by buoyancy effects compared to heat transfer effects. These observations are consistent with the conclusions that are drawn from the scaling argument presented above for the terms that contribute to the total drag on the particle. 

The third regime is the one in between these two extremes, $0.1 < Re_{BI} < 10$. This is the regime where we have an \textit{honest competition} between the two competing sources of drag modification. It is seen from the tables \ref{draggravity} and \ref{draggravity_reversed} that the drag in mixed convection in this regime can be explained within $\pm 10$\% error with linear superposition of the drag from natural and forced convection when $Re \le O(0.1)$. This is an intriguing observation and is a manifestation of the linear nature of the momentum equation (\ref{momimp}) in the low $Re$, low $Ma$ limit when the nonlinear convective Oseen-like corrections are small. 

At larger $Re$, the convective terms in the momentum equation are no longer sufficiently explained by Oseen-like corrections. This implies strong, nonlinear coupling between the governing equations leading to the breakdown of the superposition hypothesis. Cases where the linear superposition breaks down due to strong nonlinear coupling between variable fluid properties and buoyancy effects are shown in table \ref{draggravity_unsuccessful}. 

We can extend the idea of linear superposition by observing that the effect of buoyancy when $Re_{BI} < 0.1$ and the effect of heat transfer when $Re_{BI} > 10$ are negligible and linear superposition also holds well in these two regimes. The values highlighted in bold in tables \ref{draggravity} and \ref{draggravity_reversed} identify the candidates (chosen between $C_D^F$, $C_D^N$, and $C_D^F + C_D^N$) that accurately characterize the total observed drag on the sphere in mixed convection within $\pm 10$\% error. 

The success of linear superposition implies that while evaluating drag on a sphere in the low $Re$, low $Ma$ limit, it is sufficient (within $\pm10$\% error) to use the correlations developed in \cite{ganguli2019drag} for a heated sphere in forced convection and the correlations or data tables developed in literature for drag on a sphere in natural convection (for example, \cite{Churchill02}) to account for the total drag experienced by the particle. In other words, to a good approximation (within $\pm 10$ \% error), the drag modification due to heat transfer and the drag modification due to buoyancy are decoupled in this limit.

The qualitative effects of varying $Re$, $\lambda$, and $Fr$ on the features of the flow can also demonstrated by visualizing the appropriately normalized contours of the $x$-directional velocity component, $u/\sqrt{\lambda gD}$. Figure \ref{u_g_contours} shows the contours of $u/\sqrt{\lambda gD}$ for both aligned and reversed gravity setups. When $Fr > 1$, all effects are confined to a region about 15$R$ around the particle and are therefore localized effects. However, as $Re$ and $Re_{BI}$ increase, the \textit{plumelets} begin to establish a larger scale to themselves. Strong counter-rotating vortices form in the plumelet when $Fr = 0.1$ as seen in figure \ref{u_g_contours}(b). Due to significant variation in fluid density, the flow is not incompressible but has finite rate of dilatation. Contours of the appropriately scaled rate of dilatation, along with contour plots of other quantities like density, temperature, and vorticity are documented in \cite{ganguli2018computational} and are omitted in this paper for brevity. 

It is observed that as $Re_{BI}$ increases, the density and temperature variation becomes more concentrated near the particle. Figure \ref{u_g_contours} allows for a qualitative investigation into the individual manifestation of the effects of buoyancy, background uniform flow, and heat transfer from the particle in the flow features around the particle. It was demonstrated in \cite{ganguli2019drag} that while both heat transfer and convective effects increase the fore-aft asymmetry in the velocity field (in contrast to the symmetric velocity field resulting in Stokes' flow), increasing $\lambda$ causes asymmetry on the anterior stagnation side of the particle while increasing $Re$ causes asymmetry on the posterior stagnation side. This observation holds even in the presence of gravity. 

Letting figures \ref{u_g_contours}(a),(e) be the base case where $Re, \lambda, Re_{BI} = 0.1$, anterior asymmetry results when $\lambda$ in increased to 0.5 as seen in figures \ref{u_g_contours}(d),(h) while posterior asymmetry results when $Re$ is increased to 1 as seen in figures \ref{u_g_contours}(c),(g). In section \ref{scaling_viscous_pull}, it was argued that presence of gravity results in an additional force on a heated sphere due to the viscous pull caused by the migration of warm fluid near the particle in the direction opposite to gravity which scales with $Re_{BI}$. This fluid movement can be visualized when figures \ref{u_g_contours}(a),(e) and \ref{u_g_contours}(b),(f) are compared. Increasing $Re_{BI}$ from 0.1 to 1 manifests as strong vortices near the particle that contribute to increased drag.

\subsection{Test of decoupled drag contributions: Gravity perpendicular to background uniform flow}\label{perpendicular_gravity_section}
\begin{figure*}
    \centering
    \subfigure[Forced]{\includegraphics[width=0.2\textwidth]{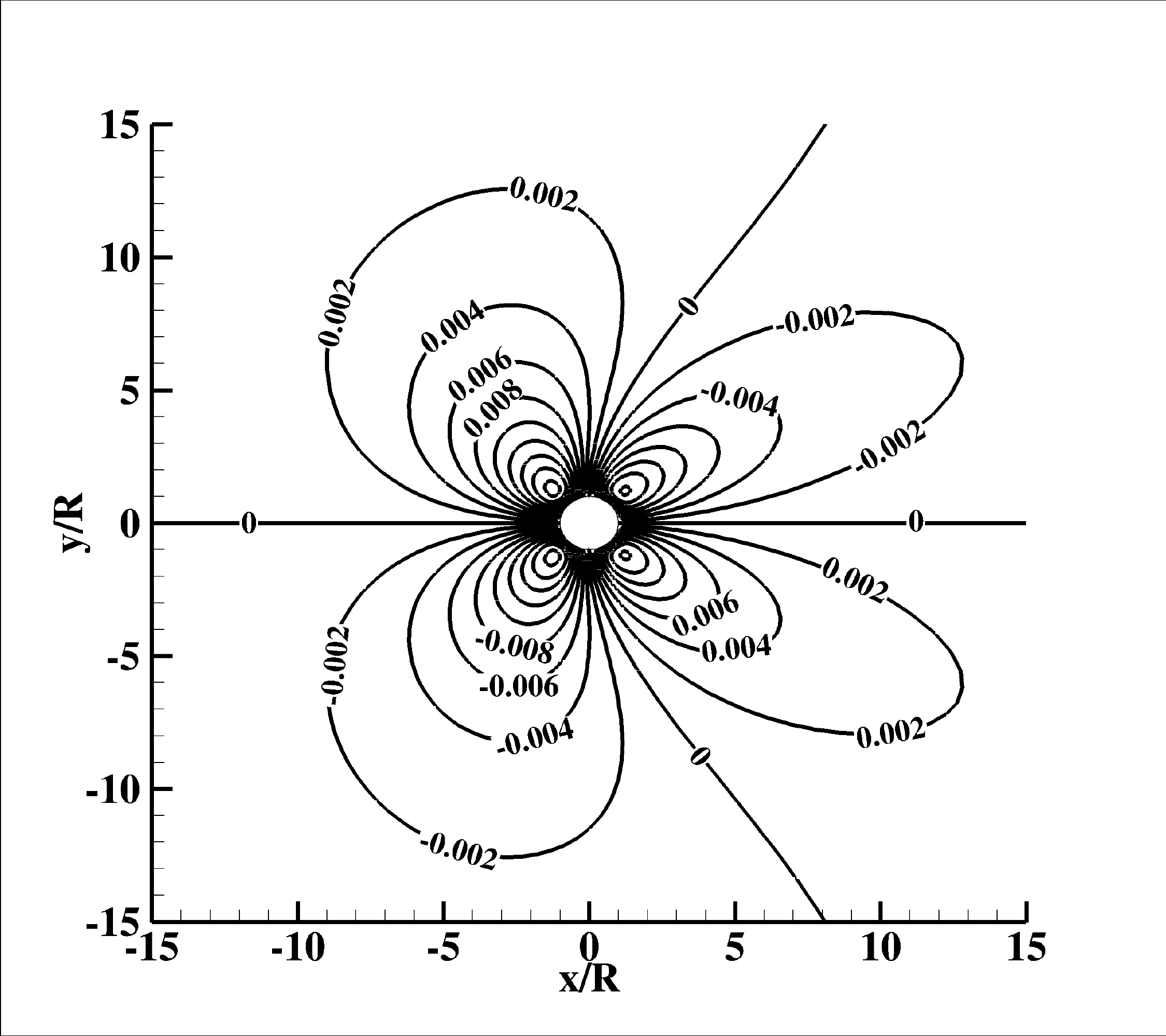}}
    \subfigure[Natural]{\includegraphics[width=0.2\textwidth]{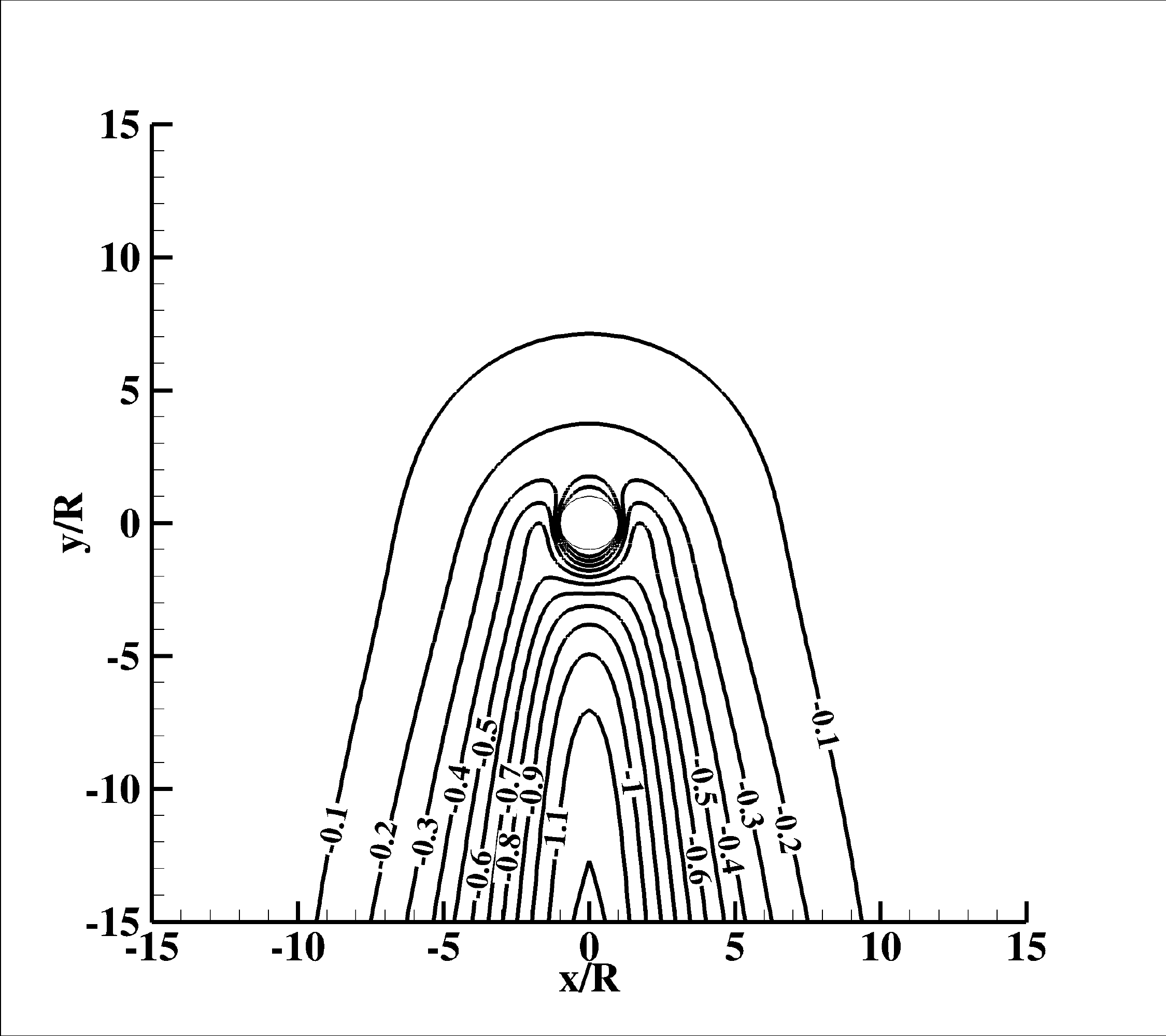}}
    \subfigure[Mixed]{\includegraphics[width=0.2\textwidth]{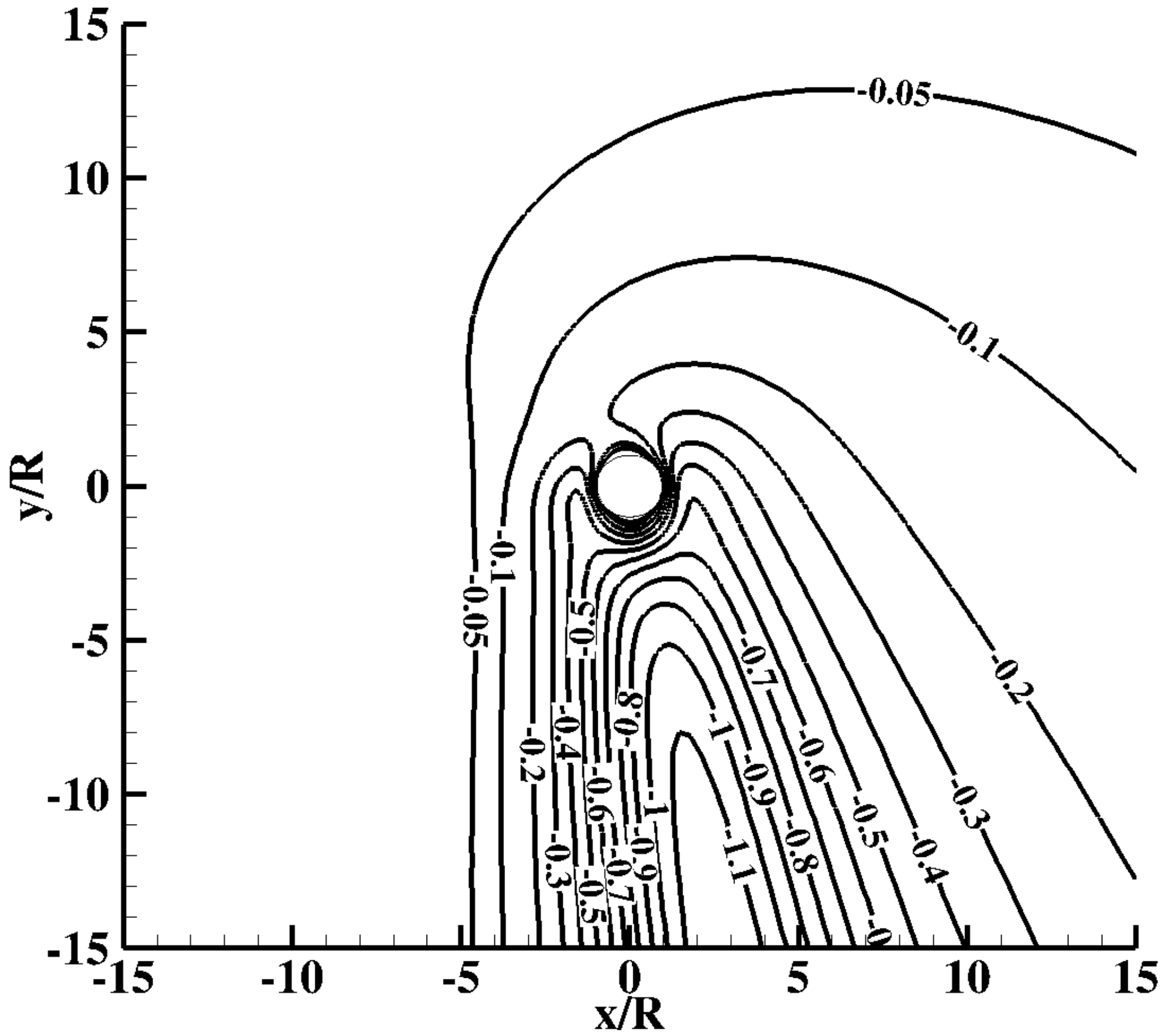}}
    \subfigure[Superposition]{\includegraphics[width=0.2\textwidth]{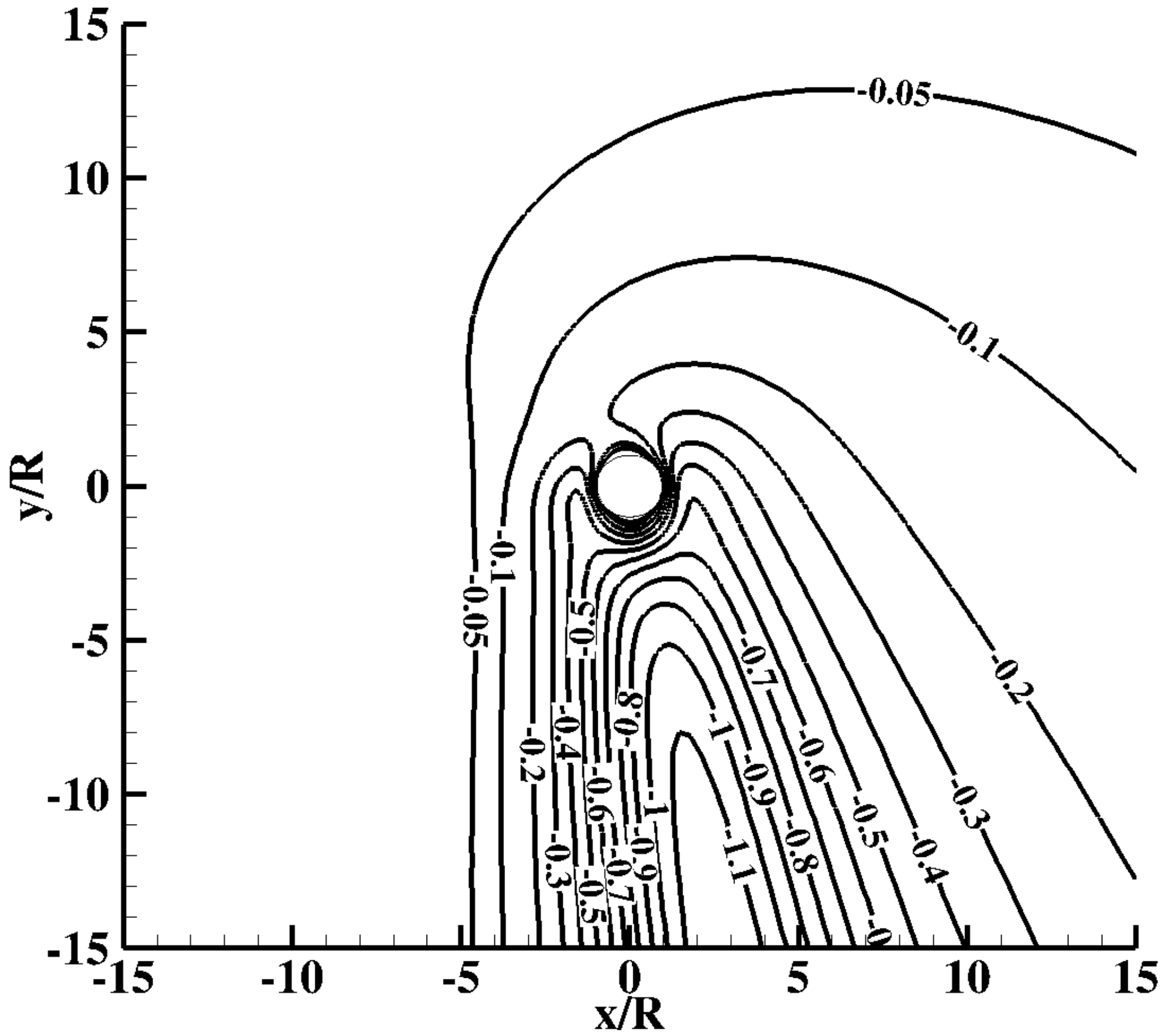}}
	\caption{Contours of $y$-directional velocity, $v/\sqrt{\lambda g D}$, for (a) forced convection at $Re = 1$ and $\lambda = 0.1$, (b) natural convection at $\lambda = 0.1$ and $Fr = 0.1$, (c) mixed convection at $Re = 1$, $\lambda = 0.1$, and $Fr = 0.1$, and (d) linear superposition of the forced and natural convection contours.\label{perp_superpose_proof}}
\end{figure*}
Encouraged by the accuracy of approximating the total drag in mixed convection as the linear superposition of the drag in forced and natural convection in the aligned and reversed gravity setups, the regime of validity of this decomposition when gravity is perpendicular to the background uniform flow is investigated in this section. This setup has also been referred to as \textit{crossflow} in literature. With reference to the coordinate axes in figure \ref{NumSetup}, uniform flow is in the $+x$ direction while gravity is in the $+y$ direction. The Froude number in this case is the ratio of the far-field convective velocity (in the $+x$ direction) and the velocity induced due to presence of buoyancy (in the $-y$ direction). 

Table \ref{draggravity_perpendicular} shows the values of the $x$ and $y$ components of the drag in mixed convection in crossflow along with the corresponding drag values in forced and natural convection. We observe from our simulations that linear superposition holds within $\pm 10$ \% error when either (i) $Re_{BI} < 0.1$, $Re_{BI} > 10$ or, (ii) $Re \ll 1$ when $0.1 < Re_{BI} < 10$. This observation is analogous to the observation in the previous subsection. 

When $Re \ll 1$, the convective terms in the momentum equation are small and negligible. These convective terms are the primary source of nonlinearities. In the absence of these terms, the momentum equation is a linear equation and coupled to the other equations only via the density. When in addition $\lambda$ is small, the contributions from buoyancy and inertia linearly add to give the resultant total force on the sphere. However, even when $Re \ll 1$ but $\lambda$ increases, non-Boussinesq effects become large and the linear approximation is no longer valid. 

To summarize, when $Re \ll 1$ and $\lambda \rightarrow 0$, the linear superposition of the drag from forced and natural convection is a good approximation (within $\pm 10$ \% error) to the total force in mixed convection even when the body force is not parallel to the direction of the far-field imposed uniform flow. Thus, in the general case when gravity is at an angle $\theta$ with the direction of the far-field uniform flow, we can decompose the buoyancy force as having a component $g \cos \theta$ in the direction parallel to the imposed far-field uniform flow and a component $g \sin \theta$ in the direction perpendicular to the flow. The former becomes a case of mixed convection while the latter becomes a case of natural convection. Under the constraints on $Re$, $Fr$, and $\lambda$ discussed above, if linear superposition holds, then the drag due to these two components can be computed separately from the correlations and linearly superposed to obtain the total drag on the spherical particle with the error margins mentioned above. 

Figure \ref{perp_superpose_proof} qualitatively demonstrates the validity of linear superposition of the flow features of the forced and free convection problems yielding the mixed convection problem where the $y$-component of the velocity has been plotted for the case when $Re = 1$, $\lambda = 0.1$, and $Fr = 0.1$. Similar plots of velocity and other quantities for all the cases in table \ref{draggravity_perpendicular} may be found in \cite{ganguli2018computational}.

\subsection{Implications for a falling heated sphere}
Given the values of the drag coefficient when gravity is anti-parallel to the far-field uniform velocity in table \ref{draggravity_reversed}, one way to measure the impact of the drag modification is to compare the terminal velocity of an unheated particle (denoted by $u_{T}^0$) falling due to gravity with that of a heated particle falling under gravity, both starting from rest. Combining the analysis in \cite{ganguli2019drag} and this paper, we can evaluate the terminal velocity taking only heat transfer effects into consideration (denoted by $u_{T}^{H}$) as well as the terminal velocity taking both heat transfer and buoyancy effects into consideration (denoted by $u_{T}^{HB}$) as the parameters $Re$, $\lambda$, and $Fr$ are varied. Note that $u_T^0$ depends only on $Re$, $u_T^H$ discussed in \cite{ganguli2019drag} depends on $Re$ and $\lambda$, and $u_T^{HB}$ incorporating buoyancy effects discussed in this paper depends on $Re$, $\lambda$, and $Fr$. 

If the fluid density is $\rho_f$, particle density is $\rho_p$, particle volume is $V$, terminal velocity of the falling particle is $u_\infty$, drag coefficient is $C_D$, acceleration due to gravity is $g$, and particle diameter is $D$, the equation \footnote{For metal particles (e.g. nickel) falling in air, $\rho_p \gg \rho_f$ so that the influence of temperature on the density ratio $\rho_f/\rho_p$ can be safely ignored for the analysis shown here. Justifiably, we can also assume zero or negligibly small contributions from fluid inertia, added mass, and, Basset-Boussinesq history terms in the Maxey-Riley-Gatignol equation. When $Re \ll 1$, added mass factor, $\beta = 3\rho_f/(\rho_f + 2\rho_p) \ll 1$.} governing the falling particle at steady state is $u_{\infty}^2 C_D = \left(1 - \rho_f/\rho_p\right)g\left(4\rho_p D/3\rho_f\right)$. If $\rho_f$, $\rho_p$, $g$, and $D$ are constant, then $\xi_H = u_{T}^{H}/u_{T}^0 = \left(C_D^0/C_D^{F}\right)^{0.5}$ and $\xi_{HB} = u_{T}^{HB}/u_{T}^0 = \left(C_D^0/C_D^M\right)^{0.5}$. For a falling particle under the influence of gravity, $C_D^M$ can be obtained from table \ref{draggravity_reversed} and the hydrostatic drag component removed to obey the governing equation stated above. 

Evaluating $\xi_H$ and $\xi_{HB}$ for the parametric variation of $Re$, $\lambda$, and $Fr$ denoted by the first 4 rows in table \ref{draggravity_reversed}, we obtain $\xi_H = 0.982, 0.983, 0.982$, and $0.914$, respectively, while $\xi_{HB} = 0.978, 0.981, 0.948$, and $0.887$, respectively. All values have been rounded off to 3 significant digits. As expected, the higher drag when heat transfer and buoyancy are present result in $u_T^0 > u_T^{H} > u_T^{HB}$. These effects are more pronounced for larger particles, smaller $Fr$, and larger values of $\lambda$.

\subsection{An aside on vorticity in the presence of gravity}
In the case of a variable density, variable property fluid, the vorticity ($\vec{\omega}$) equation can be obtained by taking the curl of the momentum equation (\ref{momentum}) which at steady state becomes
\begin{equation}\label{vorbuoy}
    \rho\left( \vec{\omega} (\nabla \cdot \vec{u}) - (\vec{\omega} \cdot \nabla)\vec{u}   \right) = \nabla \rho \times \vec{g} + \nabla \times \left(\nabla \cdot \bar{\bar{\tau}}\right)
\end{equation}
where, $\bar{\bar{\tau}} = \mu\left(\nabla \vec{u} + (\nabla \vec{u})^T - 2/3\left(\nabla \cdot \vec{u}\right)\bar{\bar{I}}\right)$ and $\bar{\bar{I}}$ is the identity tensor. For the viscous terms in equation (\ref{vorbuoy}) to be $\mathcal{O}(1)$, the inertial terms (convection, stretching, and tilting terms) scale as $Re$ while the contribution from buoyancy scales as $Re_{BV}/Re$. 

To contrast this with Stokes' flow where the particle is at the same temperature as the fluid, the vorticity equation reduces  to $\nabla^2\vec{\omega} = 0$. In the present case however, the vorticity equation has source terms owing to the contribution from buoyancy, variable density, and variable viscosity effects. Scaling analysis of equation (\ref{vorbuoy}) helps to revisit earlier observations in a new light as discussed below.

We observe from figures \ref{u_g_contours}(b) and \ref{u_g_contours}(f) that when $Re_{BI}$ is large, the migration of the warm fluid in the near vicinity of the particle surface in the direction opposite to gravity results in the formation of a \textit{plumelet} - a plume-like structure formed due to counter-rotating vortices in the presence of gravity which does not manifest as a full plume due to the low $Re$ of the flow. These counter-rotating vortical structures in the plumelets become stronger as the Froude number decreases further (which is equivalent to saying that buoyancy effects become stronger and begin to dominate). From the scaling arguments presented above, it is clear that the gravitational contribution will dominate when $Re_{BV} \gg Re \implies Re \gg Fr^2$ which is consistent with our observation that the Froude number must be small for buoyancy terms to dominate. 

Equation (\ref{vorbuoy}) also explains the validity of the low $Re$, low $Ma$, perturbation method used in \cite{ganguli2019drag} to explain the drag modification of a heated sphere in the limit $Re \rightarrow 0$ in the absence of buoyancy. The inertial terms that scale as $Re$ are small and negligible in this limit. For small but non-zero $\lambda$, equation (\ref{vorbuoy}) boils down to $\nabla^2\omega = 0$ at the first order and thus, the first order velocity correction is irrotational. The rotational part is completely contained in the Stokes-Oseen component. However, as $\lambda$ increases, the velocity correction term no longer remains irrotational and the explanation power of the perturbation model starts diminishing. When $0.01 < Re < 1$ and $\lambda$ is small, the source terms in the vorticity equation are small but non-negligible leading to the breakdown of the perturbation analysis.

% Section: Conclusions
\section{Conclusions}\label{conclusions}
A quantitative and qualitative study of the effect of heat transfer and gravity on the drag and flow features of a heated sphere in the low $Re$ and low $Ma$ regime has been presented assuming that the sphere has small $Bi$. A parametric study is performed over the governing parameters of the problem namely, $Re$, $\lambda$, $Fr$, and the orientation of gravity relative to the background uniform flow. The quantitative conclusions of this parametric study are summarized in tables \ref{draggravity}, \ref{draggravity_unsuccessful}, \ref{draggravity_reversed}, and \ref{draggravity_perpendicular}. No assumptions are made on the amount of heat addition from the sphere to the fluid or the extent of density variation that takes place in the fluid. 

Using scaling analysis and assuming low $Re$ creeping flow, the ternary parametric space can be collapsed into the buoyancy induced inertial Reynolds number, $Re_{BI}$ (or equivalently the buoyancy induced viscous Reynolds number $Re_{BV} = Re_{BI}^2$). $Re_{BV}$ is shown to be closely related to the Grashof number. Buoyancy is shown to have a significant effect on the drag experienced by the heated sphere due to the force imparted on the sphere by the migration of the warm, low density fluid in the near vicinity of the sphere in the direction opposite to gravity. Large deviations in the value of the drag coefficient (relative to that of an unheated particle) are observed when $Fr$ decreases (i.e. buoyancy effects dominate) and/or the temperature of the sphere increases. 

When the direction of gravity is aligned with the direction of the far-field uniform flow, the drag on the sphere is reduced compared to the drag in the absence of buoyancy. When the direction of gravity is anti-parallel to the direction of the far-field uniform flow, it enhances the drag on the sphere compared to the drag in absence of buoyancy. The parametric study also reveals that in the low $Re$, low $Ma$ limit, the mixed convection problem over a heated sphere in a variable property fluid can be decomposed into two simpler canonical problems whose linear superposition explains the drag on the sphere within $\pm10$\% error. These canonical problems are that of drag of a heated sphere in a uniform flow in the absence of buoyancy (forced convection) and that of drag of a heated sphere in natural convection. 

The linear superposition hypothesis not only holds for aiding and opposing flow, but also holds in crossflow. It is also shown that linear superposition extends to the flow features as well despite their large variation in the parametric space. The success of linear superposition is understood as a manifestation of the linear nature of the momentum equation in the low $Re$, low $Ma$ limit where the convective terms are small and are sufficiently modeled by Oseen-like corrections. 

Linear superposition breaks down when nonlinear coupling between variable density and buoyancy effects is significant and is characterized by the parametric space $\{\{0.1 < Re_{BI} < 10\} \cap \{Re > 0.1\}\}$. 

The quantitative implication of accounting for buoyancy effects in the drag of a heated sphere falling due to gravity is evaluated by comparing its terminal velocity with that of the terminal velocity of an unheated particle. Accounting for both heat transfer and buoyancy effects results in a lower terminal velocity than accounting only for heat transfer effects due to the enhanced drag in the former case. Similar to heat transfer effects studied in \cite{ganguli2019drag}, buoyancy effects are more pronounced for larger particles, smaller $Fr$, and larger values of $\lambda$.

% Section: Acknowledgements
\section{Acknowledgements}
We sincerely thank Prof. Howard A. Stone for his thoughtful comments on an early draft of this paper. It is also a pleasure to acknowledge insightful discussions with members of the PSAAP II team at Stanford. This work was supported by the United States Department of Energy through the Predictive Science Academic Alliance Program II (PSAAP II) at Stanford University under grant number DENA0002373-1. All numerical simulations were performed on the Certainty Cluster at the Center for Turbulence Research at Stanford University.

% Appendix
\appendix 
\section{Tables of Parametric Drag Variation}\label{tables_of_drag}
Tables \ref{draggravity}, \ref{draggravity_reversed}, and \ref{draggravity_perpendicular} list the values of the total drag on a heated sphere in mixed convection when the gravity vector and the background uniform velocity vector are parallel, anti-parallel, and orthogonal, respectively and linear superposition of forced and natural convection successfully explains the drag in mixed convection within $\pm10$\% error. The values of drag are calculated over 3 decades of variation in both $Re$ and $Fr$. Table \ref{draggravity_unsuccessful} lists cases where linear superposition fails to explain the mixed convection drag within $\pm10$\% error in the aligned gravity setup.

% Table of drag when gravity is present: Aligned Gravity, Successful Linear Superposition
\begin{table*}[!htbp]
    \begin{center}
        \centering
        \begin{tabular}{p{1cm} p{1cm} | p{0.7cm} p{0.7cm} p{0.7cm} | p{1.5cm} p{1.5cm} p{1.5cm} | p{2cm} p{2cm} p{3cm} p{0.01cm}}
            \centering{$Re_{BI}$} & \centering{$Re_{BV}$} & \centering{$Re$} & \centering{$\lambda$} & \centering{$Fr$}                 & \centering{$C_D^F$} & \centering{$C_D^N$}  & \centering{$C_D^M$}  & \centering{$100(C_D^F/C_D^M)$} & \centering{$100(C_D^N/C_D^M)$} & \centering{$100((C_D^F + C_D^N)/C_D^M)$} & \\
            \hline
            \centering{0.01}      & \centering{0.0001}    & \centering{0.1}  & \centering{0.1}       & \centering{\multirow{2}{*}{10}}  & \centering{253.09}  & \centering{-0.36}    & \centering{254.00}   & \centering{\textbf{99.64}}     & \centering{-0.14}              & \centering{\textbf{99.50}}  & \\
            \centering{0.1}       & \centering{0.01}      & \centering{1}    & \centering{0.1}       &                                  & \centering{28.12}   & \centering{-0.32}    & \centering{28.21}    & \centering{\textbf{99.69}}     & \centering{-1.13}              & \centering{\textbf{98.56}}  & \\
            \hline 
            \centering{0.1}       & \centering{0.01}      & \centering{0.1}  & \centering{0.1}       & \centering{\multirow{3}{*}{1}}   & \centering{253.09}  & \centering{-40.74}   & \centering{213.50}   & \centering{118.54}             & \centering{-19.08}             & \centering{\textbf{99.46}}  & \\
            \centering{0.1}       & \centering{0.01}      & \centering{0.1}  & \centering{0.5}       &                                  & \centering{292.34}  & \centering{-20.37}   & \centering{277.10}   & \centering{105.50}             & \centering{-7.35}              & \centering{\textbf{98.15}}  & \\
            \centering{0.1}       & \centering{0.01}      & \centering{0.1}  & \centering{1}         &                                  & \centering{341.99}  & \centering{-22.92}   & \centering{327.99}   & \centering{104.27}             & \centering{-6.99}              & \centering{\textbf{97.28}}  & \\
            \hline 
            \centering{1}         & \centering{1}         & \centering{0.1}  & \centering{0.1}       & \centering{\multirow{8}{*}{0.1}} & \centering{253.09}  & \centering{-2850.27} & \centering{-2613.38} & \centering{-9.68}              & \centering{109.06}             & \centering{\textbf{99.38}}  & \\
            \centering{1}         & \centering{1}         & \centering{0.1}  & \centering{0.5}       &                                  & \centering{292.34}  & \centering{-1844.41} & \centering{-1564.27} & \centering{-18.69}             & \centering{117.91}             & \centering{\textbf{99.22}}  & \\
            \centering{1}         & \centering{1}         & \centering{0.1}  & \centering{1}         &                                  & \centering{341.99}  & \centering{-1870.39} & \centering{-1551.83} & \centering{-22.04}             & \centering{120.53}             & \centering{\textbf{98.49}}  & \\
            \centering{10}        & \centering{100}       & \centering{1}    & \centering{0.1}       &                                  & \centering{28.12}   & \centering{-1720.40} & \centering{-1744.08} & \centering{-1.61}              & \centering{\textbf{98.64}}     & \centering{\textbf{97.03}}  & \\
            \centering{10}        & \centering{100}       & \centering{1}    & \centering{0.5}       &                                  & \centering{32.66}   & \centering{-676.34}  & \centering{-676.63}  & \centering{-4.83}              & \centering{\textbf{99.96}}     & \centering{\textbf{95.13}}  & \\
            \centering{10}        & \centering{100}       & \centering{1}    & \centering{1}         &                                  & \centering{37.84}   & \centering{-529.92}  & \centering{-526.12}  & \centering{-7.19}              & \centering{\textbf{100.72}}    & \centering{\textbf{93.53}}  & \\
            \centering{100}       & \centering{10000}     & \centering{10}   & \centering{0.1}       &                                  & \centering{4.38}    & \centering{-1420.91} & \centering{-1421.08} & \centering{-0.31}              & \centering{\textbf{99.99}}     & \centering{\textbf{99.68}}  & \\
            \centering{100}       & \centering{10000}     & \centering{10}   & \centering{0.5}       &                                  & \centering{4.87}    & \centering{-354.55}  & \centering{-356.13}  & \centering{-1.37}              & \centering{\textbf{99.56}}     & \centering{\textbf{98.19}}  & 
        \end{tabular}
        \caption{Cases where linear superposition of drag on a sphere in forced and natural convection successfully explains the drag in mixed convection within $\pm$10\% error when gravity is parallel to the background uniform flow. \label{draggravity}}
    \end{center}
\end{table*}

% Table of drag when gravity is present: Aligned Gravity, Unsuccessful Linear Superposition
\begin{table*}[!htbp]
    \begin{center}
        \centering
        \begin{tabular}{p{1cm} p{1cm} | p{0.7cm} p{0.7cm} p{0.7cm} | p{1.5cm} p{1.5cm} p{1.5cm} | p{2cm} p{2cm} p{3cm} p{0.01cm}}
            \centering{$Re_{BI}$} & \centering{$Re_{BV}$} & \centering{$Re$} & \centering{$\lambda$} & \centering{$Fr$}               & \centering{$C_D^F$} & \centering{$C_D^N$}  & \centering{$C_D^M$}  & \centering{$100(C_D^F/C_D^M)$} & \centering{$100(C_D^N/C_D^M)$} & \centering{$100((C_D^F + C_D^N)/C_D^M)$} & \\
            \hline 
            \centering{1}         & \centering{1}         & \centering{1}    & \centering{0.1}       & \centering{\multirow{3}{*}{1}} & \centering{28.12}   & \centering{-29.75}   & \centering{1.97}     & \centering{1428.77}            & \centering{-1511.59}           & \centering{-82.82} & \\
            \centering{1}         & \centering{1}         & \centering{1}    & \centering{0.5}       &                                & \centering{32.66}   & \centering{-20.60}   & \centering{15.53}    & \centering{210.37}             & \centering{-132.69}            & \centering{77.68}  & \\
            \centering{1}         & \centering{1}         & \centering{1}    & \centering{1}         &                                & \centering{37.84}   & \centering{-23.06}   & \centering{19.63}    & \centering{192.73}             & \centering{-117.45}            & \centering{75.28}  & \\
        \end{tabular}
        \caption{Cases where linear superposition of drag on a sphere in forced and natural convection fails to explain the drag in mixed convection within $\pm$10\% error when gravity is parallel to the background uniform flow. \label{draggravity_unsuccessful}}
    \end{center}
\end{table*}

% Table of drag when gravity is present: Reversed Gravity, Successful Linear Superposition
\begin{table*}[!htbp]
    \begin{center}
        \centering
        \begin{tabular}{p{1cm} p{1cm} | p{0.7cm} p{0.7cm} p{0.7cm} | p{1.5cm} p{1.5cm} p{1.5cm} | p{2cm} p{2cm} p{3cm} p{0.01cm}}
            \centering{$Re_{BI}$} & \centering{$Re_{BV}$} & \centering{$Re$} & \centering{$\lambda$} & \centering{$Fr$}                 & \centering{$C_D^F$} & \centering{$C_D^N$}  & \centering{$C_D^M$}  & \centering{$100(C_D^F/C_D^M)$} & \centering{$100(C_D^N/C_D^M)$} & \centering{$100((C_D^F + C_D^N)/C_D^M)$} & \\
            \hline 
            \centering{0.01}      & \centering{0.0001}    & \centering{0.1}  & \centering{0.1}       & \centering{\multirow{2}{*}{10}}  & \centering{253.09}  & \centering{0.36}     & \centering{254.29}   & \centering{\textbf{99.53}}     & \centering{0.14}               & \centering{\textbf{99.67}} & \\
            \centering{0.1}       & \centering{0.01}      & \centering{1}    & \centering{0.1}       &                                  & \centering{28.12}   & \centering{0.32}     & \centering{28.59}    & \centering{\textbf{98.37}}     & \centering{1.12}               & \centering{\textbf{99.49}} & \\
            \hline
            \centering{0.1}       & \centering{0.01}      & \centering{0.1}  & \centering{0.1}       & \centering{\multirow{2}{*}{1}}   & \centering{253.09}  & \centering{40.74}    & \centering{296.95}   & \centering{85.23}              & \centering{13.72}              & \centering{\textbf{98.95}} & \\
            \centering{0.1}       & \centering{0.01}      & \centering{0.1}  & \centering{0.5}       &                                  & \centering{292.34}  & \centering{20.37}    & \centering{316.48}   & \centering{92.37}              & \centering{6.44}               & \centering{\textbf{98.81}} & \\
            \hline
            \centering{1}         & \centering{1}         & \centering{0.1}  & \centering{0.1}       & \centering{\multirow{3}{*}{0.1}} & \centering{253.09}  & \centering{2850.27}  & \centering{3138.83}  & \centering{8.06}               & \centering{\textbf{90.81}}     & \centering{\textbf{98.87}} & \\
            \centering{10}        & \centering{100}       & \centering{1}    & \centering{0.1}       &                                  & \centering{28.12}   & \centering{1720.40}  & \centering{1784.57}  & \centering{1.58}               & \centering{\textbf{96.40}}     & \centering{\textbf{97.98}} & \\
            \centering{100}       & \centering{10000}     & \centering{10}   & \centering{0.1}       &                                  & \centering{4.38}    & \centering{1420.91}  & \centering{1468.61}  & \centering{0.30}               & \centering{\textbf{96.75}}     & \centering{\textbf{97.05}} & 
        \end{tabular}
        \caption{Cases where linear superposition of drag on a sphere in forced and natural convection successfully explains the drag in mixed convection within $\pm$10\% error when gravity is anti-parallel to the background uniform flow. \label{draggravity_reversed}}
    \end{center}
\end{table*}

% Table of drag when gravity is present: Perpendicular Gravity, Successful Linear Superposition
\begin{table*}[!htbp]
    \begin{center}
        \centering
        \begin{tabular}{p{1cm} p{1cm} | p{0.8cm} p{0.8cm} p{0.8cm} | p{1.6cm} p{1.6cm} p{1.6cm} p{1.6cm} | p{2.5cm} p{0.01cm}}
            \centering{$Re_{BI}$} & \centering{$Re_{BV}$} & \centering{$Re$} & \centering{$\lambda$} & \centering{$Fr$}                 & \centering{$C_D^F$}  & \centering{$C_D^N$}  & \centering{$C_D^x$}  & \centering{$C_D^y$}  & \centering{$100\sqrt{\frac{(C_D^F)^2+(C_D^N)^2}{(C_D^x)^2+(C_D^y)^2}}$} & \\ 
            \hline
            \centering{0.01}      & \centering{0.0001}    & \centering{0.1}  & \centering{0.1}       & \centering{\multirow{2}{*}{10}}  & \centering{253.09}   & \centering{-0.36}    & \centering{253.27}   & \centering{-0.31}    & \centering{\textbf{99.93}} & \\
            \centering{0.1}       & \centering{0.01}      & \centering{1}    & \centering{0.1}       &                                  & \centering{28.12}    & \centering{-0.32}    & \centering{28.48}    & \centering{-0.28}    & \centering{\textbf{98.75}} & \\
            \hline 
            \centering{0.1}       & \centering{0.01}      & \centering{0.1}  & \centering{0.1}       & \centering{\multirow{2}{*}{1}}   & \centering{253.09}   & \centering{-40.74}   & \centering{257.40}   & \centering{-29.80}   & \centering{\textbf{98.93}} & \\
            \centering{1}         & \centering{1}         & \centering{1}    & \centering{0.1}       &                                  & \centering{28.12}    & \centering{-29.75}   & \centering{41.77}    & \centering{-21.22}   & \centering{87.37}          & \\
            \hline
            \centering{1}         & \centering{1}         & \centering{0.1}  & \centering{0.1}       & \centering{\multirow{2}{*}{0.1}} & \centering{253.09}   & \centering{-2850.27} & \centering{319.91}   & \centering{-2846.14} & \centering{\textbf{99.91}} & \\
            \centering{10}        & \centering{100}       & \centering{1}    & \centering{0.1}       &                                  & \centering{28.12}    & \centering{-1720.40} & \centering{177.09}   & \centering{-1715.48} & \centering{\textbf{99.77}} &
        \end{tabular}
        \caption{Cases where linear superposition of drag on a sphere in forced and natural convection successfully explains the drag in mixed convection within $\pm$10\% error when gravity is perpendicular to the background uniform flow.}
        \label{draggravity_perpendicular}
    \end{center}
\end{table*}

% Bibliography
\bibliography{SwetavaGanguli_PaperIII_References}

\providecommand{\noopsort}[1]{}\providecommand{\singleletter}[1]{#1}%
\begin{thebibliography}{50}%
\makeatletter
\providecommand \@ifxundefined [1]{%
 \@ifx{#1\undefined}
}%
\providecommand \@ifnum [1]{%
 \ifnum #1\expandafter \@firstoftwo
 \else \expandafter \@secondoftwo
 \fi
}%
\providecommand \@ifx [1]{%
 \ifx #1\expandafter \@firstoftwo
 \else \expandafter \@secondoftwo
 \fi
}%
\providecommand \natexlab [1]{#1}%
\providecommand \enquote  [1]{``#1''}%
\providecommand \bibnamefont  [1]{#1}%
\providecommand \bibfnamefont [1]{#1}%
\providecommand \citenamefont [1]{#1}%
\providecommand \href@noop [0]{\@secondoftwo}%
\providecommand \href [0]{\begingroup \@sanitize@url \@href}%
\providecommand \@href[1]{\@@startlink{#1}\@@href}%
\providecommand \@@href[1]{\endgroup#1\@@endlink}%
\providecommand \@sanitize@url [0]{\catcode `\\12\catcode `\$12\catcode
  `\&12\catcode `\#12\catcode `\^12\catcode `\_12\catcode `\%12\relax}%
\providecommand \@@startlink[1]{}%
\providecommand \@@endlink[0]{}%
\providecommand \url  [0]{\begingroup\@sanitize@url \@url }%
\providecommand \@url [1]{\endgroup\@href {#1}{\urlprefix }}%
\providecommand \urlprefix  [0]{URL }%
\providecommand \Eprint [0]{\href }%
\providecommand \doibase [0]{https://doi.org/}%
\providecommand \selectlanguage [0]{\@gobble}%
\providecommand \bibinfo  [0]{\@secondoftwo}%
\providecommand \bibfield  [0]{\@secondoftwo}%
\providecommand \translation [1]{[#1]}%
\providecommand \BibitemOpen [0]{}%
\providecommand \bibitemStop [0]{}%
\providecommand \bibitemNoStop [0]{.\EOS\space}%
\providecommand \EOS [0]{\spacefactor3000\relax}%
\providecommand \BibitemShut  [1]{\csname bibitem#1\endcsname}%
\let\auto@bib@innerbib\@empty
%</preamble>
\bibitem [{\citenamefont {Ganguli}\ and\ \citenamefont
  {Lele}(2019{\natexlab{a}})}]{ganguli2019drag}%
  \BibitemOpen
  \bibfield  {author} {\bibinfo {author} {\bibfnamefont {S.}~\bibnamefont
  {Ganguli}}\ and\ \bibinfo {author} {\bibfnamefont {S.~K.}\ \bibnamefont
  {Lele}},\ }\bibfield  {title} {\bibinfo {title} {Drag of a heated sphere at
  low reynolds numbers in the absence of buoyancy},\ }\href@noop {} {\bibfield
  {journal} {\bibinfo  {journal} {J. Fluid Mech.}\ }\textbf {\bibinfo {volume}
  {869}},\ \bibinfo {pages} {264} (\bibinfo {year}
  {2019}{\natexlab{a}})}\BibitemShut {NoStop}%
\bibitem [{\citenamefont {Turner}(1973)}]{Turner73}%
  \BibitemOpen
  \bibfield  {author} {\bibinfo {author} {\bibfnamefont {J.~S.}\ \bibnamefont
  {Turner}},\ }\href@noop {} {\emph {\bibinfo {title} {{Buoyancy effects in
  fluids}}}}\ (\bibinfo  {publisher} {Cambridge University Press, Cambridge,
  England},\ \bibinfo {year} {1973})\BibitemShut {NoStop}%
\bibitem [{\citenamefont {Chandrasekhar}(1961)}]{Chandrasekhar61}%
  \BibitemOpen
  \bibfield  {author} {\bibinfo {author} {\bibfnamefont {S.}~\bibnamefont
  {Chandrasekhar}},\ }\href@noop {} {\emph {\bibinfo {title} {{Hydrodynamic and
  hydromagnetic stability}}}}\ (\bibinfo  {publisher} {Oxford University Press,
  Oxford, England},\ \bibinfo {year} {1961})\BibitemShut {NoStop}%
\bibitem [{\citenamefont {Yih}(1960)}]{Yih60}%
  \BibitemOpen
  \bibfield  {author} {\bibinfo {author} {\bibfnamefont {C.~S.}\ \bibnamefont
  {Yih}},\ }\bibfield  {title} {\bibinfo {title} {{Exact solutions for steady
  two-dimensional flow of a stratified fluid}},\ }\href@noop {} {\bibfield
  {journal} {\bibinfo  {journal} {J. Fluid Mech.}\ }\textbf {\bibinfo {volume}
  {9}},\ \bibinfo {pages} {161} (\bibinfo {year} {1960})}\BibitemShut {NoStop}%
\bibitem [{\citenamefont {Drazin}(1969)}]{Drazin69}%
  \BibitemOpen
  \bibfield  {author} {\bibinfo {author} {\bibfnamefont {P.~G.}\ \bibnamefont
  {Drazin}},\ }\bibfield  {title} {\bibinfo {title} {{Nonlinear internal
  gravity waves in a slightly stratified atmosphere}},\ }\href@noop {}
  {\bibfield  {journal} {\bibinfo  {journal} {J. Fluid Mech.}\ }\textbf
  {\bibinfo {volume} {36}},\ \bibinfo {pages} {433} (\bibinfo {year}
  {1969})}\BibitemShut {NoStop}%
\bibitem [{\citenamefont {Batchelor}(1954)}]{Batchelor54}%
  \BibitemOpen
  \bibfield  {author} {\bibinfo {author} {\bibfnamefont {G.~K.}\ \bibnamefont
  {Batchelor}},\ }\bibfield  {title} {\bibinfo {title} {{Heat convection and
  buoyancy effects in fluids}},\ }\href@noop {} {\bibfield  {journal} {\bibinfo
   {journal} {Quart. J. Roy. Met. Soc}\ }\textbf {\bibinfo {volume} {80}},\
  \bibinfo {pages} {339} (\bibinfo {year} {1954})}\BibitemShut {NoStop}%
\bibitem [{\citenamefont {Townsend}(1970)}]{Townsend70}%
  \BibitemOpen
  \bibfield  {author} {\bibinfo {author} {\bibfnamefont {A.~A.}\ \bibnamefont
  {Townsend}},\ }\bibfield  {title} {\bibinfo {title} {{Entrainment and the
  structure of turbulent flow}},\ }\href@noop {} {\bibfield  {journal}
  {\bibinfo  {journal} {J. Fluid Mech.}\ }\textbf {\bibinfo {volume} {41}},\
  \bibinfo {pages} {13} (\bibinfo {year} {1970})}\BibitemShut {NoStop}%
\bibitem [{\citenamefont {Lagerstrom}(1988)}]{Lagerstrom88}%
  \BibitemOpen
  \bibfield  {author} {\bibinfo {author} {\bibfnamefont {P.~A.}\ \bibnamefont
  {Lagerstrom}},\ }\href@noop {} {\emph {\bibinfo {title} {{Matched asymptotic
  expansions - ideas and techniques}}}},\ \bibinfo {series} {Applied
  Mathematical Series}, Vol.~\bibinfo {volume} {76}\ (\bibinfo  {publisher}
  {Springer-Verlag},\ \bibinfo {address} {New York},\ \bibinfo {year}
  {1988})\BibitemShut {NoStop}%
\bibitem [{\citenamefont {Merk}\ and\ \citenamefont {Prims}(1953)}]{Merk53}%
  \BibitemOpen
  \bibfield  {author} {\bibinfo {author} {\bibfnamefont {H.~J.}\ \bibnamefont
  {Merk}}\ and\ \bibinfo {author} {\bibfnamefont {J.~A.}\ \bibnamefont
  {Prims}},\ }\bibfield  {title} {\bibinfo {title} {{Thermal convection in
  laminar boundary layers. I}},\ }\href@noop {} {\bibfield  {journal} {\bibinfo
   {journal} {Appl. Sci. Res., Series A}\ }\textbf {\bibinfo {volume} {4}},\
  \bibinfo {pages} {11} (\bibinfo {year} {1953})}\BibitemShut {NoStop}%
\bibitem [{\citenamefont {Acrivos}(1960{\natexlab{a}})}]{Acrivos60nn}%
  \BibitemOpen
  \bibfield  {author} {\bibinfo {author} {\bibfnamefont {A.}~\bibnamefont
  {Acrivos}},\ }\bibfield  {title} {\bibinfo {title} {{A theoretical analysis
  of laminar natural convection heat transfer to non-Newtonian fluids}},\
  }\href@noop {} {\bibfield  {journal} {\bibinfo  {journal} {AIChE J.}\
  }\textbf {\bibinfo {volume} {6}},\ \bibinfo {pages} {584} (\bibinfo {year}
  {1960}{\natexlab{a}})}\BibitemShut {NoStop}%
\bibitem [{\citenamefont {Potter}\ and\ \citenamefont
  {Riley}(1980)}]{Potter80}%
  \BibitemOpen
  \bibfield  {author} {\bibinfo {author} {\bibfnamefont {J.~M.}\ \bibnamefont
  {Potter}}\ and\ \bibinfo {author} {\bibfnamefont {N.}~\bibnamefont {Riley}},\
  }\bibfield  {title} {\bibinfo {title} {{Free convection from a heated sphere
  at large Grashof number}},\ }\href@noop {} {\bibfield  {journal} {\bibinfo
  {journal} {J. Fluid Mech.}\ }\textbf {\bibinfo {volume} {100}},\ \bibinfo
  {pages} {769} (\bibinfo {year} {1980})}\BibitemShut {NoStop}%
\bibitem [{\citenamefont {Fendell}(1968)}]{Fendell68}%
  \BibitemOpen
  \bibfield  {author} {\bibinfo {author} {\bibfnamefont {F.~E.}\ \bibnamefont
  {Fendell}},\ }\bibfield  {title} {\bibinfo {title} {{Laminar natural
  convection about an isothermally heated sphere at small Grashof number}},\
  }\href@noop {} {\bibfield  {journal} {\bibinfo  {journal} {J. Fluid Mech.}\
  }\textbf {\bibinfo {volume} {34}},\ \bibinfo {pages} {163} (\bibinfo {year}
  {1968})}\BibitemShut {NoStop}%
\bibitem [{\citenamefont {Hieber}\ and\ \citenamefont
  {Gebhart}(1969)}]{Hieber69}%
  \BibitemOpen
  \bibfield  {author} {\bibinfo {author} {\bibfnamefont {C.~A.}\ \bibnamefont
  {Hieber}}\ and\ \bibinfo {author} {\bibfnamefont {B.}~\bibnamefont
  {Gebhart}},\ }\bibfield  {title} {\bibinfo {title} {{Mixed convection from a
  sphere at small Reynolds and Grashof numbers}},\ }\href@noop {} {\bibfield
  {journal} {\bibinfo  {journal} {J. Fluid Mech.}\ }\textbf {\bibinfo {volume}
  {38}},\ \bibinfo {pages} {137} (\bibinfo {year} {1969})}\BibitemShut
  {NoStop}%
\bibitem [{\citenamefont {Hossain}\ and\ \citenamefont
  {Gebhart}(1970)}]{Hossain70}%
  \BibitemOpen
  \bibfield  {author} {\bibinfo {author} {\bibfnamefont {M.~A.}\ \bibnamefont
  {Hossain}}\ and\ \bibinfo {author} {\bibfnamefont {B.}~\bibnamefont
  {Gebhart}},\ }\bibfield  {title} {\bibinfo {title} {{Natural convection about
  a sphere at low Grashof number}},\ }in\ \href@noop {} {\emph {\bibinfo
  {booktitle} {Fourth International Heat Transfer Conference, Versailles,
  Paris. NCl. 6. A.I.Ch.E, New York}}},\ Vol.~\bibinfo {volume} {5}\ (\bibinfo
  {year} {1970})\BibitemShut {NoStop}%
\bibitem [{\citenamefont {Mathers}\ \emph {et~al.}(1957)\citenamefont
  {Mathers}, \citenamefont {Madden},\ and\ \citenamefont {Piret}}]{Mathers57}%
  \BibitemOpen
  \bibfield  {author} {\bibinfo {author} {\bibfnamefont {W.~G.}\ \bibnamefont
  {Mathers}}, \bibinfo {author} {\bibfnamefont {A.~J.}\ \bibnamefont
  {Madden}},\ and\ \bibinfo {author} {\bibfnamefont {E.~L.}\ \bibnamefont
  {Piret}},\ }\bibfield  {title} {\bibinfo {title} {{Simultaneous heat and mass
  transfer in free convection}},\ }\href@noop {} {\bibfield  {journal}
  {\bibinfo  {journal} {Ind. Engng. Chem.}\ }\textbf {\bibinfo {volume} {49}},\
  \bibinfo {pages} {961} (\bibinfo {year} {1957})}\BibitemShut {NoStop}%
\bibitem [{\citenamefont {Tsubouchi}\ and\ \citenamefont
  {Sato}(1960)}]{Tsubouchi60}%
  \BibitemOpen
  \bibfield  {author} {\bibinfo {author} {\bibfnamefont {T.}~\bibnamefont
  {Tsubouchi}}\ and\ \bibinfo {author} {\bibfnamefont {S.}~\bibnamefont
  {Sato}},\ }\bibfield  {title} {\bibinfo {title} {{Heat transfer from fine
  wires and particles by natural convection}},\ }\href@noop {} {\bibfield
  {journal} {\bibinfo  {journal} {Res. Inst. High Speed Mech., Tohoku Univ.}\
  }\textbf {\bibinfo {volume} {12}},\ \bibinfo {pages} {127} (\bibinfo {year}
  {1960})}\BibitemShut {NoStop}%
\bibitem [{\citenamefont {Yuge}(1960)}]{Yuge60}%
  \BibitemOpen
  \bibfield  {author} {\bibinfo {author} {\bibfnamefont {T.}~\bibnamefont
  {Yuge}},\ }\bibfield  {title} {\bibinfo {title} {{Experiments on heat
  transfer from sphere including combined natural and forced convection}},\
  }\href@noop {} {\bibfield  {journal} {\bibinfo  {journal} {Trans. ASME}\
  }\textbf {\bibinfo {volume} {C82}},\ \bibinfo {pages} {214} (\bibinfo {year}
  {1960})}\BibitemShut {NoStop}%
\bibitem [{\citenamefont {Spiegel}\ and\ \citenamefont
  {Veronis}(1960)}]{Spiegel60}%
  \BibitemOpen
  \bibfield  {author} {\bibinfo {author} {\bibfnamefont {E.~A.}\ \bibnamefont
  {Spiegel}}\ and\ \bibinfo {author} {\bibfnamefont {G.}~\bibnamefont
  {Veronis}},\ }\bibfield  {title} {\bibinfo {title} {{On the Boussinesq
  approximation for a compressible fluid}},\ }\href@noop {} {\bibfield
  {journal} {\bibinfo  {journal} {J. Astrophys.}\ }\textbf {\bibinfo {volume}
  {131}},\ \bibinfo {pages} {442} (\bibinfo {year} {1960})}\BibitemShut
  {NoStop}%
\bibitem [{\citenamefont {Spjut}(1985)}]{Spjut85}%
  \BibitemOpen
  \bibfield  {author} {\bibinfo {author} {\bibfnamefont {R.~E.}\ \bibnamefont
  {Spjut}},\ }\emph {\bibinfo {title} {{Heat transfer to and position control
  of electrodynamically suspended micron-sized particles}}},\ \href@noop {}
  {Ph.D. thesis} (\bibinfo {year} {1985})\BibitemShut {NoStop}%
\bibitem [{\citenamefont {Geoola}\ and\ \citenamefont
  {Cornish}(1981)}]{Geoola81}%
  \BibitemOpen
  \bibfield  {author} {\bibinfo {author} {\bibfnamefont {F.}~\bibnamefont
  {Geoola}}\ and\ \bibinfo {author} {\bibfnamefont {A.~R.~H.}\ \bibnamefont
  {Cornish}},\ }\bibfield  {title} {\bibinfo {title} {{Numerical solution of
  steady-state free convective heat transfer from a solid sphere}},\
  }\href@noop {} {\bibfield  {journal} {\bibinfo  {journal} {Int. J. Heat Mass
  Transfer}\ }\textbf {\bibinfo {volume} {24}},\ \bibinfo {pages} {1369}
  (\bibinfo {year} {1981})}\BibitemShut {NoStop}%
\bibitem [{\citenamefont {Geoola}\ and\ \citenamefont
  {Cornish}(1982)}]{Geoola82}%
  \BibitemOpen
  \bibfield  {author} {\bibinfo {author} {\bibfnamefont {F.}~\bibnamefont
  {Geoola}}\ and\ \bibinfo {author} {\bibfnamefont {A.~R.~H.}\ \bibnamefont
  {Cornish}},\ }\bibfield  {title} {\bibinfo {title} {{Numerical simulation of
  free convective heat transfer from a sphere}},\ }\href@noop {} {\bibfield
  {journal} {\bibinfo  {journal} {Int. J. Heat Mass Transfer}\ }\textbf
  {\bibinfo {volume} {25}},\ \bibinfo {pages} {1677} (\bibinfo {year}
  {1982})}\BibitemShut {NoStop}%
\bibitem [{\citenamefont {Jia}\ and\ \citenamefont {Gogos}(1996)}]{Jia96}%
  \BibitemOpen
  \bibfield  {author} {\bibinfo {author} {\bibfnamefont {H.}~\bibnamefont
  {Jia}}\ and\ \bibinfo {author} {\bibfnamefont {G.}~\bibnamefont {Gogos}},\
  }\bibfield  {title} {\bibinfo {title} {{Laminar natural convection heat
  transfer from isothermal spheres}},\ }\href@noop {} {\bibfield  {journal}
  {\bibinfo  {journal} {Int. J. Heat Mass Transfer}\ }\textbf {\bibinfo
  {volume} {39}},\ \bibinfo {pages} {1603} (\bibinfo {year}
  {1996})}\BibitemShut {NoStop}%
\bibitem [{\citenamefont {Dudek}\ \emph {et~al.}(1988)\citenamefont {Dudek},
  \citenamefont {Fletcher}, \citenamefont {Longwell},\ and\ \citenamefont
  {Sarofim}}]{Dudek88}%
  \BibitemOpen
  \bibfield  {author} {\bibinfo {author} {\bibfnamefont {D.~R.}\ \bibnamefont
  {Dudek}}, \bibinfo {author} {\bibfnamefont {T.~H.}\ \bibnamefont {Fletcher}},
  \bibinfo {author} {\bibfnamefont {J.~P.}\ \bibnamefont {Longwell}},\ and\
  \bibinfo {author} {\bibfnamefont {A.~F.}\ \bibnamefont {Sarofim}},\
  }\bibfield  {title} {\bibinfo {title} {{Natural convection induced drag
  forces on spheres at low Grashof numbers: Comparison of theory with
  experiment}},\ }\href@noop {} {\bibfield  {journal} {\bibinfo  {journal}
  {Int. J. Heat Mass Transfer}\ }\textbf {\bibinfo {volume} {31}},\ \bibinfo
  {pages} {863} (\bibinfo {year} {1988})}\BibitemShut {NoStop}%
\bibitem [{\citenamefont {Schenkels}\ and\ \citenamefont
  {Schenk}(1969)}]{Schenkels69}%
  \BibitemOpen
  \bibfield  {author} {\bibinfo {author} {\bibfnamefont {F.~A.~M.}\
  \bibnamefont {Schenkels}}\ and\ \bibinfo {author} {\bibfnamefont
  {J.}~\bibnamefont {Schenk}},\ }\bibfield  {title} {\bibinfo {title}
  {{Dissolution of solid spheres by isothermal free convection}},\ }\href@noop
  {} {\bibfield  {journal} {\bibinfo  {journal} {J. Chem. Eng. Sci.}\ }\textbf
  {\bibinfo {volume} {24}},\ \bibinfo {pages} {585} (\bibinfo {year}
  {1969})}\BibitemShut {NoStop}%
\bibitem [{\citenamefont {Jaluria}\ and\ \citenamefont
  {Gebhart}(1975)}]{Jaluria75}%
  \BibitemOpen
  \bibfield  {author} {\bibinfo {author} {\bibfnamefont {Y.}~\bibnamefont
  {Jaluria}}\ and\ \bibinfo {author} {\bibfnamefont {B.}~\bibnamefont
  {Gebhart}},\ }\bibfield  {title} {\bibinfo {title} {{On the buoyancy-induced
  flow arising from a heated hemisphere}},\ }\href@noop {} {\bibfield
  {journal} {\bibinfo  {journal} {Int. J. Heat Mass Transfer}\ }\textbf
  {\bibinfo {volume} {18}},\ \bibinfo {pages} {415} (\bibinfo {year}
  {1975})}\BibitemShut {NoStop}%
\bibitem [{\citenamefont {Churchill}(1990)}]{Churchill02}%
  \BibitemOpen
  \bibfield  {author} {\bibinfo {author} {\bibfnamefont {S.~W.}\ \bibnamefont
  {Churchill}},\ }\bibfield  {title} {\bibinfo {title} {{Free convection around
  immersed bodies}},\ }in\ \href@noop {} {\emph {\bibinfo {booktitle}
  {{Hemisphere handbook of heat exchanger design}}}},\ \bibinfo {editor}
  {edited by\ \bibinfo {editor} {\bibfnamefont {G.~F.}\ \bibnamefont
  {Hewitt}}}\ (\bibinfo  {publisher} {Hemisphere Publishing Corporation, Taylor
  and Francis Group},\ \bibinfo {year} {1990})\BibitemShut {NoStop}%
\bibitem [{\citenamefont {Clift}\ \emph {et~al.}(1978)\citenamefont {Clift},
  \citenamefont {Grace},\ and\ \citenamefont {Weber}}]{CGW78}%
  \BibitemOpen
  \bibfield  {author} {\bibinfo {author} {\bibfnamefont {R.}~\bibnamefont
  {Clift}}, \bibinfo {author} {\bibfnamefont {J.}~\bibnamefont {Grace}},\ and\
  \bibinfo {author} {\bibfnamefont {M.~E.}\ \bibnamefont {Weber}},\ }\href@noop
  {} {\emph {\bibinfo {title} {{Bubbles, drops and particles}}}}\ (\bibinfo
  {publisher} {Dover Publications},\ \bibinfo {year} {1978})\BibitemShut
  {NoStop}%
\bibitem [{\citenamefont {Proudman}\ and\ \citenamefont
  {Pearson}(1957)}]{Proudman57}%
  \BibitemOpen
  \bibfield  {author} {\bibinfo {author} {\bibfnamefont {I.}~\bibnamefont
  {Proudman}}\ and\ \bibinfo {author} {\bibfnamefont {J.~R.~A.}\ \bibnamefont
  {Pearson}},\ }\bibfield  {title} {\bibinfo {title} {{Expansions at small
  {R}eynolds numbers for the flow past a sphere and a circular cylinder}},\
  }\href@noop {} {\bibfield  {journal} {\bibinfo  {journal} {J.~Fluid Mech.}\
  }\textbf {\bibinfo {volume} {2}},\ \bibinfo {pages} {237} (\bibinfo {year}
  {1957})}\BibitemShut {NoStop}%
\bibitem [{\citenamefont {Acrivos}(1960{\natexlab{b}})}]{Acrivos60}%
  \BibitemOpen
  \bibfield  {author} {\bibinfo {author} {\bibfnamefont {A.}~\bibnamefont
  {Acrivos}},\ }\bibfield  {title} {\bibinfo {title} {{Momentum and heat
  transfer in laminar boundary-layer flows of non-Newtonian fluids past
  external surfaces}},\ }\href@noop {} {\bibfield  {journal} {\bibinfo
  {journal} {AIChE J.}\ }\textbf {\bibinfo {volume} {6}},\ \bibinfo {pages}
  {584} (\bibinfo {year} {1960}{\natexlab{b}})}\BibitemShut {NoStop}%
\bibitem [{\citenamefont {Stewart}(1971)}]{Stewart71}%
  \BibitemOpen
  \bibfield  {author} {\bibinfo {author} {\bibfnamefont {W.~E.}\ \bibnamefont
  {Stewart}},\ }\bibfield  {title} {\bibinfo {title} {{Asymptotic calculation
  of free convection in laminar three-dimensional systems}},\ }\href@noop {}
  {\bibfield  {journal} {\bibinfo  {journal} {Int. J. Heat Mass Transfer}\
  }\textbf {\bibinfo {volume} {14}},\ \bibinfo {pages} {1013} (\bibinfo {year}
  {1971})}\BibitemShut {NoStop}%
\bibitem [{\citenamefont {Chen}\ and\ \citenamefont {Mucoglu}(1978)}]{Chen78}%
  \BibitemOpen
  \bibfield  {author} {\bibinfo {author} {\bibfnamefont {T.~S.}\ \bibnamefont
  {Chen}}\ and\ \bibinfo {author} {\bibfnamefont {A.}~\bibnamefont {Mucoglu}},\
  }\bibfield  {title} {\bibinfo {title} {{Mixed convection about a sphere with
  uniform surface heat flux}},\ }\href@noop {} {\bibfield  {journal} {\bibinfo
  {journal} {J. Heat Mass Transfer}\ }\textbf {\bibinfo {volume} {115:21}},\
  \bibinfo {pages} {542} (\bibinfo {year} {1978})}\BibitemShut {NoStop}%
\bibitem [{\citenamefont {Woo}(1971)}]{Woo71}%
  \BibitemOpen
  \bibfield  {author} {\bibinfo {author} {\bibfnamefont {S.~W.}\ \bibnamefont
  {Woo}},\ }\emph {\bibinfo {title} {{Simultaneous free and forced convection
  around submerged cylinders and spheres}}},\ \href@noop {} {Ph.D. thesis},\
  \bibinfo  {school} {{McMaster University, Hamilton, Ontario}} (\bibinfo
  {year} {1971})\BibitemShut {NoStop}%
\bibitem [{\citenamefont {Acrivos}(1958)}]{Acrivos58}%
  \BibitemOpen
  \bibfield  {author} {\bibinfo {author} {\bibfnamefont {A.}~\bibnamefont
  {Acrivos}},\ }\bibfield  {title} {\bibinfo {title} {{Combined laminar free-
  and forced-convection heat transfer in external flows}},\ }\href@noop {}
  {\bibfield  {journal} {\bibinfo  {journal} {AIChE J.}\ }\textbf {\bibinfo
  {volume} {4}},\ \bibinfo {pages} {285} (\bibinfo {year} {1958})}\BibitemShut
  {NoStop}%
\bibitem [{\citenamefont {Pearson}\ and\ \citenamefont
  {Dickson}(1968)}]{Pearson68}%
  \BibitemOpen
  \bibfield  {author} {\bibinfo {author} {\bibfnamefont {R.~S.}\ \bibnamefont
  {Pearson}}\ and\ \bibinfo {author} {\bibfnamefont {P.~F.}\ \bibnamefont
  {Dickson}},\ }\bibfield  {title} {\bibinfo {title} {{Free convective effects
  on Stokes flow mass transfer}},\ }\href@noop {} {\bibfield  {journal}
  {\bibinfo  {journal} {AIChE J.}\ }\textbf {\bibinfo {volume} {14}},\ \bibinfo
  {pages} {903} (\bibinfo {year} {1968})}\BibitemShut {NoStop}%
\bibitem [{\citenamefont {Hatton}\ \emph {et~al.}(1970)\citenamefont {Hatton},
  \citenamefont {James},\ and\ \citenamefont {Swire}}]{Hatton70}%
  \BibitemOpen
  \bibfield  {author} {\bibinfo {author} {\bibfnamefont {A.~P.}\ \bibnamefont
  {Hatton}}, \bibinfo {author} {\bibfnamefont {D.~D.}\ \bibnamefont {James}},\
  and\ \bibinfo {author} {\bibfnamefont {H.~W.}\ \bibnamefont {Swire}},\
  }\bibfield  {title} {\bibinfo {title} {{Combined forced and natural
  convection with low-speed air flow over horizontal cylinders}},\ }\href@noop
  {} {\bibfield  {journal} {\bibinfo  {journal} {J. Fluid Mech.}\ }\textbf
  {\bibinfo {volume} {42}},\ \bibinfo {pages} {17} (\bibinfo {year}
  {1970})}\BibitemShut {NoStop}%
\bibitem [{\citenamefont {Oosthuizen}\ and\ \citenamefont
  {Madan}(1971)}]{Oosthuizen71}%
  \BibitemOpen
  \bibfield  {author} {\bibinfo {author} {\bibfnamefont {P.~H.}\ \bibnamefont
  {Oosthuizen}}\ and\ \bibinfo {author} {\bibfnamefont {S.~J.}\ \bibnamefont
  {Madan}},\ }\bibfield  {title} {\bibinfo {title} {{The effect of flow
  direction on combined convective heat transfer from cylinders to air}},\
  }\href@noop {} {\bibfield  {journal} {\bibinfo  {journal} {J. Heat Transfer}\
  }\textbf {\bibinfo {volume} {93}},\ \bibinfo {pages} {240} (\bibinfo {year}
  {1971})}\BibitemShut {NoStop}%
\bibitem [{\citenamefont {Mograbi}\ and\ \citenamefont
  {Bar-Ziv}(2005{\natexlab{a}})}]{Mograbi05a}%
  \BibitemOpen
  \bibfield  {author} {\bibinfo {author} {\bibfnamefont {E.}~\bibnamefont
  {Mograbi}}\ and\ \bibinfo {author} {\bibfnamefont {E.}~\bibnamefont
  {Bar-Ziv}},\ }\bibfield  {title} {\bibinfo {title} {{Dynamics of a spherical
  particle in mixed convection flow field}},\ }\href@noop {} {\bibfield
  {journal} {\bibinfo  {journal} {Aerosol Science}\ }\textbf {\bibinfo {volume}
  {36}},\ \bibinfo {pages} {387} (\bibinfo {year}
  {2005}{\natexlab{a}})}\BibitemShut {NoStop}%
\bibitem [{\citenamefont {Mograbi}\ and\ \citenamefont
  {Bar-Ziv}(2005{\natexlab{b}})}]{Mograbi05b}%
  \BibitemOpen
  \bibfield  {author} {\bibinfo {author} {\bibfnamefont {E.}~\bibnamefont
  {Mograbi}}\ and\ \bibinfo {author} {\bibfnamefont {E.}~\bibnamefont
  {Bar-Ziv}},\ }\bibfield  {title} {\bibinfo {title} {{On the mixed convection
  hydrodynamic force on a sphere}},\ }\href@noop {} {\bibfield  {journal}
  {\bibinfo  {journal} {Aerosol Science}\ }\textbf {\bibinfo {volume} {36}},\
  \bibinfo {pages} {1177} (\bibinfo {year} {2005}{\natexlab{b}})}\BibitemShut
  {NoStop}%
\bibitem [{\citenamefont {Kotouc}\ \emph
  {et~al.}(2009{\natexlab{a}})\citenamefont {Kotouc}, \citenamefont {Bouchet},\
  and\ \citenamefont {Dusek}}]{Kotouc09a}%
  \BibitemOpen
  \bibfield  {author} {\bibinfo {author} {\bibfnamefont {M.}~\bibnamefont
  {Kotouc}}, \bibinfo {author} {\bibfnamefont {G.}~\bibnamefont {Bouchet}},\
  and\ \bibinfo {author} {\bibfnamefont {J.}~\bibnamefont {Dusek}},\ }\bibfield
   {title} {\bibinfo {title} {{Drag and flow reversal in mixed convection past
  a heated sphere}},\ }\href@noop {} {\bibfield  {journal} {\bibinfo  {journal}
  {Phys. of Fluids}\ }\textbf {\bibinfo {volume} {21}},\ \bibinfo {pages} {14}
  (\bibinfo {year} {2009}{\natexlab{a}})}\BibitemShut {NoStop}%
\bibitem [{\citenamefont {Kotouc}\ \emph
  {et~al.}(2009{\natexlab{b}})\citenamefont {Kotouc}, \citenamefont {Bouchet},\
  and\ \citenamefont {Dusek}}]{Kotouc09b}%
  \BibitemOpen
  \bibfield  {author} {\bibinfo {author} {\bibfnamefont {M.}~\bibnamefont
  {Kotouc}}, \bibinfo {author} {\bibfnamefont {G.}~\bibnamefont {Bouchet}},\
  and\ \bibinfo {author} {\bibfnamefont {J.}~\bibnamefont {Dusek}},\ }\bibfield
   {title} {\bibinfo {title} {{Transition to turbulence in the wake of a fixed
  sphere in mixed convection}},\ }\href@noop {} {\bibfield  {journal} {\bibinfo
   {journal} {J. Fluid Mech.}\ }\textbf {\bibinfo {volume} {625}},\ \bibinfo
  {pages} {205} (\bibinfo {year} {2009}{\natexlab{b}})}\BibitemShut {NoStop}%
\bibitem [{\citenamefont {Nirmalkar}\ and\ \citenamefont
  {Chhabra}(2013)}]{Nirmalkar13}%
  \BibitemOpen
  \bibfield  {author} {\bibinfo {author} {\bibfnamefont {N.}~\bibnamefont
  {Nirmalkar}}\ and\ \bibinfo {author} {\bibfnamefont {R.~P.}\ \bibnamefont
  {Chhabra}},\ }\bibfield  {title} {\bibinfo {title} {{Mixed convection from a
  heated sphere in power-law fluids}},\ }\href@noop {} {\bibfield  {journal}
  {\bibinfo  {journal} {Chem. Engg. Sci.}\ }\textbf {\bibinfo {volume} {89}},\
  \bibinfo {pages} {49} (\bibinfo {year} {2013})}\BibitemShut {NoStop}%
\bibitem [{\citenamefont {Ganguli}\ and\ \citenamefont
  {Lele}(2019{\natexlab{b}})}]{ganguli2019low}%
  \BibitemOpen
  \bibfield  {author} {\bibinfo {author} {\bibfnamefont {S.}~\bibnamefont
  {Ganguli}}\ and\ \bibinfo {author} {\bibfnamefont {S.~K.}\ \bibnamefont
  {Lele}},\ }\bibfield  {title} {\bibinfo {title} {Low mach, compressibility,
  and finite size effects of localized uniform heat sources in a gas},\
  }\href@noop {} {\bibfield  {journal} {\bibinfo  {journal} {Theoretical and
  Computational Fluid Dynamics}\ }\textbf {\bibinfo {volume} {33}},\ \bibinfo
  {pages} {341} (\bibinfo {year} {2019}{\natexlab{b}})}\BibitemShut {NoStop}%
\bibitem [{\citenamefont {Panton}(2005)}]{Panton}%
  \BibitemOpen
  \bibfield  {author} {\bibinfo {author} {\bibfnamefont {R.~L.}\ \bibnamefont
  {Panton}},\ }\href@noop {} {\emph {\bibinfo {title} {{Incompressible
  {f}low}}}},\ \bibinfo {edition} {3rd}\ ed.\ (\bibinfo  {publisher} {John
  Wiley \& Sons, Inc.},\ \bibinfo {year} {2005})\BibitemShut {NoStop}%
\bibitem [{Note1()}]{Note1}%
  \BibitemOpen
  \bibinfo {note} {For air, $\kappa = 2 \times 10^{-2}W/mK$, $T_\infty = 300K$.
  For particle, $D=1\mu m$, emissivity, $\epsilon = 1$, Nusselt Number, $Nu=2$,
  $T_p=1200K$. $\sigma _{SB}$ is the Stefan-Boltzmann constant. Convective
  thermal power, $P_c = \kappa A Nu (T_p - T_\infty )/D$. Radiative thermal
  power, $P_r = \sigma _{SB} A (T_p^4 - T_{\infty }^4)$. Then $P_r/P_c =
  \protect \mathcal {O}(10^{-3})$.}\BibitemShut {Stop}%
\bibitem [{\citenamefont {Ham}(2007)}]{Ham07}%
  \BibitemOpen
  \bibfield  {author} {\bibinfo {author} {\bibfnamefont {F.}~\bibnamefont
  {Ham}},\ }\bibfield  {title} {\bibinfo {title} {{An efficient scheme for
  large eddy simulation of low-{Ma} combustion in complex configurations}},\
  }\href@noop {} {\bibfield  {journal} {\bibinfo  {journal} {Center for
  Turbulence Research, Annual Res. Briefs}\ } (\bibinfo {year}
  {2007})}\BibitemShut {NoStop}%
\bibitem [{\citenamefont {Kundu}\ \emph {et~al.}(2014)\citenamefont {Kundu},
  \citenamefont {Cohen},\ and\ \citenamefont {Dowling}}]{Kundu51}%
  \BibitemOpen
  \bibfield  {author} {\bibinfo {author} {\bibfnamefont {P.~K.}\ \bibnamefont
  {Kundu}}, \bibinfo {author} {\bibfnamefont {I.~M.}\ \bibnamefont {Cohen}},\
  and\ \bibinfo {author} {\bibfnamefont {D.~R.}\ \bibnamefont {Dowling}},\
  }\href@noop {} {\emph {\bibinfo {title} {{Fluid mechanics}}}}\ (\bibinfo
  {publisher} {Academic Press},\ \bibinfo {year} {2014})\BibitemShut {NoStop}%
\bibitem [{\citenamefont {Vincenti}\ and\ \citenamefont
  {Kruger}(1965)}]{Vincenti65}%
  \BibitemOpen
  \bibfield  {author} {\bibinfo {author} {\bibfnamefont {W.~G.}\ \bibnamefont
  {Vincenti}}\ and\ \bibinfo {author} {\bibfnamefont {C.~H.}\ \bibnamefont
  {Kruger}},\ }\href@noop {} {\emph {\bibinfo {title} {{Introduction to
  physical gas dynamics}}}}\ (\bibinfo  {publisher} {John Wiley and Sons,
  Inc.},\ \bibinfo {year} {1965})\BibitemShut {NoStop}%
\bibitem [{\citenamefont {Ganguli}(2018)}]{ganguli2018computational}%
  \BibitemOpen
  \bibfield  {author} {\bibinfo {author} {\bibfnamefont {S.}~\bibnamefont
  {Ganguli}},\ }\emph {\bibinfo {title} {Computational analysis of canonical
  problems arising in the interaction of heated particles and a fluid}},\
  \href@noop {} {Ph.D. thesis},\ \bibinfo  {school} {Stanford University}
  (\bibinfo {year} {2018})\BibitemShut {NoStop}%
\bibitem [{\citenamefont {Lagerstrom}(1964)}]{Lagerstrom64}%
  \BibitemOpen
  \bibfield  {author} {\bibinfo {author} {\bibfnamefont {P.~A.}\ \bibnamefont
  {Lagerstrom}},\ }\href@noop {} {\emph {\bibinfo {title} {{Laminar flow
  theory}}}}\ (\bibinfo  {publisher} {Princeton University Press, Princeton,
  New Jersey},\ \bibinfo {year} {1964})\BibitemShut {NoStop}%
\bibitem [{Note2()}]{Note2}%
  \BibitemOpen
  \bibinfo {note} {For metal particles (e.g. nickel) falling in air, $\rho _p
  \gg \rho _f$ so that the influence of temperature on the density ratio $\rho
  _f/\rho _p$ can be safely ignored for the analysis shown here. Justifiably,
  we can also assume zero or negligibly small contributions from fluid inertia,
  added mass, and, Basset-Boussinesq history terms in the Maxey-Riley-Gatignol
  equation. When $Re \ll 1$, added mass factor, $\beta = 3\rho _f/(\rho _f +
  2\rho _p) \ll 1$.}\BibitemShut {Stop}%
\end{thebibliography}%

% End of Document
\end{document}